# Ultrafast Nano-Imaging and Optical Control of Hyperbolic Phonon Polaritons at hBN/$WS_2$ Heterojunctions

*Kazuki Kamada[1,2], Keisuke Shinokita[1,3], Fanyu Zeng[4], Ryo Kitaura[4], Kenji Watanabe[5], Takashi Taniguchi[4], Alexander Paarmann[6], Masahiro Shibuta[1,2], Takashi Kumagai[1,3*], Jun Nishida[1,3*]*

[1]Institute for Molecular Science, National Institutes of Natural Sciences, Okazaki, Aichi, 444-8585, Japan

[2]Department of Physics and Electronics, Graduate School of Engineering, Osaka Metropolitan University, 1-1, Gakuen-cho, Naka-ku, Sakai, Osaka 599-8531, Japan

[3]The Graduate University for Advanced Studies, SOKENDAI, Hayama 240-0193, Kanagawa, Japan

[4]Research Center for Materials Nanoarchitectonics, National Institute for Materials Science, 1-1 Namiki, Tsukuba 305-0044, Japan

[5]Research Center for Electronic and Optical Materials, National Institute for Materials Science, 1-1 Namiki, Tsukuba 305-0044, Japan

[6]Fritz-Haber-Institut der Max-Planck-Gesellschaft, Faradayweg 4-6, 14195 Berlin, Germany

ABSTRACT

Manipulating nanoscale light–matter interactions on ultrafast time scales is indispensable for future polaritonic devices. Hyperbolic phonon polaritons (HPhPs) in van der Waals materials enable deep subwavelength confinement of electromagnetic fields in the infrared region and long-distance propagation of polaritonic waves. However, achieving ultrafast imaging and optical control of HPhPs remains a major challenge. Here, we demonstrate the direct observation of transient modulation of HPhPs induced by local photocarrier generation in $WS_2$/hBN heterostructures using ultrafast infrared scanning near-field optical microscopy. We implement grating-based spectral filtering of broadband near-field scattering to simultaneously achieve nanoscale and femtosecond spatiotemporal resolution together with fine spectral selectivity. This ultrafast nano-imaging technique reveals that photocarriers in $WS_2$ modulate the polaritonic field amplitudes and wavelengths of HPhPs in hBN. Theoretical simulations corroborate that these changes arise from photoinduced changes in $WS_2$ dielectric properties. This approach offers a versatile platform for exploring ultrafast polaritonic dynamics at the nanoscale.

The ability to control light at deep-subwavelength scales is central to nanophotonics. Polaritons, which are hybrid modes of light and collective material excitations, provide a means to confine a light field to subwavelength scale. In the infrared (IR) region, electromagnetic fields couple to optical phonons in polar crystals to form phonon polaritons, whose surface modes enable strong field confinement in mid- to far-IR applications.[1]

Anisotropic polar crystals support hyperbolic phonon polaritons (HPhPs), in which the dielectric permittivity tensor components possess opposite signs within the Reststrahlen band and give rise to hyperbolic dispersion.[2] HPhPs propagate as volume-confined modes with large momenta and extremely short wavelengths in the IR region, enabling deep-subwavelength optics,[3–5] ultrasensitive molecular spectroscopy,[6] and interfacial thermal management.[7] While hyperbolic materials were initially explored as metamaterials made of engineered nanostructures,[8] HPhPs are now known to occur intrinsically in anisotropic crystals such as hBN,[2–4,9–12] $\alpha$-$MoO_3$,[13,14] $\alpha$-$V_2O_5$[15] and many other materials with a high degree of anisotropy.[16–26] These natural hyperbolic materials offer low-loss phonon resonances and broadband hyperbolic responses without requiring complex nanofabrication.[27] In the past decade, HPhPs have been directly visualized in real space using infrared scattering-type scanning near-field optical microscopy (IR *s*-SNOM), elucidating a plethora of fascinating polaritonic phenomena.[28,29]

Because HPhPs are sensitive to the surrounding dielectric environment, their amplitudes and wavelengths can be tailored by embedding the polaritonic materials in a variety of heterostructures.[30] Particularly, in van der Waals (vdW) materials, the polaritonic properties can be tuned by substrates,[12,31–38] vdW heterostructures,[39–41] and coupling with molecular vibrations.[42,43] In addition, active control has been achieved using phase-change materials,[44–47] electrostatic gates,[48–52] and the application of magnetic fields,[53] laying foundations for polaritonic devices.[54]

In addition to static control, polaritons can be perturbed on ultrafast timescales using optical excitation, enabling prospects for ultrafast polaritonic devices. Ultrafast *s*-SNOM has successfully visualized transient polaritonic phenomena in various materials,[55–61] leveraging its ability to image polaritons as spatially oscillating waves[2,62,63] and to capture their femtosecond perturbations.

However, applying ultrafast IR *s*-SNOM imaging to HPhPs is challenging because of their strong and spectrally confined dispersion; broadband excitation launches propagating polaritons with a wide range of wavelengths, leading to a rapid wave packet dephasing and preventing clear real-space observation of polaritonic waves. This is not an issue for weakly dispersive plasmon polaritons,[56] hybrid plasmon-phonon polaritons,[61] or exciton polaritons,[64] for which ultrafast polaritonic nano-imaging has been successfully performed. Yet, the same scheme is not applicable to HPhPs which are strongly dispersive within a narrow Reststrahlen band, and a new approach is necessary to visualize the ultrafast control of HPhPs in real space.

Here, we demonstrate ultrafast control of HPhPs in $WS_2$/hBN heterostructures, visualized by ultrafast IR *s*-SNOM with simultaneous spatial, temporal, and spectral resolutions. Narrowband detection enabled by diffraction-grating filtering achieves a spectral resolution of ~10 $cm^{-1}$ while maintaining a temporal resolution of ~150 fs. Upon femtosecond visible excitation, photocarriers generated in $WS_2$ transiently modulate HPhPs in an adjacent hBN layer (Figure 1). These transient responses manifest as interference fringes that are spatially confined to hBN regions covered by thin $WS_2$ layers and persist only over the photocarrier lifetime. Furthermore, in heterostructures where hBN is stacked atop bulk $WS_2$, we directly visualize ultrafast modulation of the polaritonic wavelength in real space by exploiting the enhanced interaction volume between hBN and $WS_2$.

Numerical simulations confirm that these effects originate from photo-induced changes in the dielectric function of $WS_2$. Our approach enables femtosecond optical control of HPhPs and near-field nano-imaging of strongly dispersive polaritons, opening a pathway toward actively tunable ultrafast polaritonic devices.

Figure 2a illustrates our experimental configuration, and we provide further details in the Supporting Information Note S1. Ultrafast nano-imaging was implemented by a phase-resolved pump–probe scheme.[65] Near-field IR signals were detected by lock-in detection of the MCT output at the third harmonic of the tip-tapping frequency ($3\omega_{AFM}$) for static (ground state) imaging, denoted as $S_3$. The transient signal in pump–probe measurements was obtained by sideband demodulation at $2\omega_{AFM} \pm \Omega_{chop}$, denoted as $\Delta S_2$, which is sufficiently spatially localized with a high signal-to-noise ratio,[66,67] enabling selective and sensitive detection of the pump-induced near-field response that reflects changes in the local dielectric response.[65–67] To eliminate potential background contributions to the near-field signals, the tip-scattered mid-IR field was interfered with a reference pulse and detected in the two-phase homodyne scheme.[68] As a unique feature of the current setup, a diffraction grating was placed in front of the MCT detector, achieving an effective detection bandwidth (spectral resolution) of ~10 $cm^{-1}$. The center frequency within the broadband probe spectrum was tuned by azimuthally rotating the grating.

We first examined ground-state near-field images of HPhPs in a bare hBN layer to verify that the grating-coupled detection in broadband IR *s*-SNOM yields results consistent with those obtained under conventional continuous-wave (CW) mid-IR probe.[2] Figure 2b displays the AFM topography of an hBN flake on an $SiO_2$ substrate and the corresponding *s*-SNOM images ($S_3$) acquired with the grating-coupled detection scheme at 1477 $cm^{-1}$ and 1517 $cm^{-1}$. Despite the use of the broadband mid-IR probe, the interference fringes characteristic of HPhPs were clearly observed near the hBN edge, with spatial periods varying with the selected mid-IR frequency. In contrast, such fringes were absent under direct broadband detection without a grating (Supporting Information Figure S1).

In this ground-state *s*-SNOM imaging, the AFM tip acts as a nanoscale antenna that provides the large in-plane momentum required to couple free-space light to highly confined phonon polaritons. Previous CW *s*-SNOM measurements have revealed that the fringes predominantly arise from the interference between tip-launched and edge-reflected phonon polaritons.[2,9,31,69] This interference generates fringes in the *s*-SNOM images with a period corresponding to half the polaritonic wavelength ($\lambda_p/2$), based on which we converted the observed polariton fringe spacing to the polaritonic wavelength. In some studies, edge-launched polaritons were also observed far away from the hBN edge, exhibiting fringe periods identical to the polaritonic wavelength $\lambda_p$.[42,69] We also confirmed the presence of these edge-launched polariton fringes in some of our measurements (Supporting Information Note S2, Figure S2).

The results in Figure 2b validate that grating-coupled ultrafast IR *s*-SNOM can accurately characterize strongly dispersive HPhPs in hBN. Notably, the observed polaritonic wavelengths are strongly dependent on the IR frequency, reflecting strong dispersion of the HPhP (Figure 2b, middle and right). We performed IR frequency dependent nano-imaging to retrieve a dispersion relationship (Figure 2c). The measured dispersion agrees quantitatively with the peaks in the imaginary part of the Fresnel reflection coefficient for p-polarized light (Im $r_p$),[2,70] which is calculated by transfer-matrix formalism for an air/hBN/$SiO_2$ structure based on a measured hBN thickness of 147 nm and previously reported dielectric functions.[9,71] These results further corroborate that the broadband near-field scattering can be spectrally filtered to isolate narrowband components, enabling the reconstruction of quasi-monochromatic polariton waves.

We note that another approach to image strongly dispersive polaritons based on a mid-IR pulse is to limit the spectral bandwidth incident to the whole asymmetric Michelson interferometer,[55,72] but this method inevitably sacrifices temporal resolution. In contrast, our grating-coupled detection scheme offers an advantage enabling frequency-resolved nano-imaging with femtosecond temporal resolution (Supporting Information Figure S3), thereby overcoming the inherent temporal–frequency trade-off.

Next, we demonstrate the visualization of ultrafast HPhP dynamics in $WS_2$/hBN heterostructures. Figure 3a shows the AFM image of a heterostructure fabricated by chemical vapor deposition (CVD), in which $WS_2$ layers were directly grown over exfoliated hBN flakes on an $SiO_2$ substrate.[73] The AFM topography confirms the presence of triangular $WS_2$ domains near the hBN edge, and the thickness of $WS_2$ is estimated to be approximately 1.5 nm, corresponding to a bilayer. Figure 3b presents ground-state near-field images at 1498 $cm^{-1}$ (without a visible pump), where polariton interference fringes are visible near the hBN edge.

We found that, upon visible-pump excitation, the transient mid-IR near-field images ($\Delta S_2$) are dominated by a characteristic fringe that appears in the region of the $WS_2$ layer and decays on the picosecond timescale (Figure 3c). The fringe period depends sensitively on the frequency of the infrared near-field probe, similar to ground-state imaging of hBN, indicating that the observed fringes originate from the strongly dispersive HPhPs of hBN (Supporting Information Figure S4). Similar transient fringe patterns were also observed in other $WS_2$ flakes, confirming the reproducibility of the phenomenon (Supporting Information Figure S5).

To quantify the effect of $WS_2$ excitation on HPhPs in hBN, we analyzed line profiles from near-field images. The $\Delta S_2$ profiles exhibited anti-phase oscillations with nearly the same periods as the $S_3$ fringe patterns, indicating that the photo-induced change arises primarily from ultrafast suppression of polaritonic field strength (Figure 3d). The relaxation of the transient appearance of the HPhPs induced by the visible-pump excitation occurs on a timescale comparable to that of photocarriers generated in CVD-grown multilayer $WS_2$,[67] as confirmed by near-field pump–probe traces (Supporting Information Figure S3). Therefore, the transient fringe patterns are associated with photocarriers in the $WS_2$ layer that locally modulate the dielectric response. We emphasize that our grating-coupled polariton imaging is particularly advantageous for visualizing local ultrafast phenomena in inhomogeneous structures such as CVD-grown $WS_2$ layers on hBN.

The above results for the CVD-grown $WS_2$/hBN heterostructure suggest that photocarriers generated adjacent to hBN can dynamically modulate the amplitudes of HPhPs through a change in the local dielectric function. However, small interaction volumes of the deposited $WS_2$ limit more dramatic modulation of polariton properties, such as polaritonic wavelength.[58] To enhance the interaction volume between hBN and $WS_2$, we then examined another heterostructure in which an hBN flake is dry-transferred to a relatively thick (bulk) $WS_2$ flake. This configuration allows injection of a larger number of photocarriers compared to the atomically thin $WS_2$ layers, resulting in more pronounced modulation of HPhPs of hBN. Figure 4a shows AFM topography of hBN/$WS_2$/$SiO_2$, where the thickness of hBN and $WS_2$ are determined to be 38 nm and 376 nm, respectively. Figure 4b and 4c display the ground-state and transient-state near-field images at 1425 $cm^{-1}$ at $\Delta t = 0$ ps, respectively. The fringe profiles of $S_3$ and $\Delta S_2$ along the direction indicated by the white dashed arrow in Figure 4b, plotted as a function of the distance from the hBN edge ($L$), are shown in Figure 4d. To evaluate the polaritonic wavelength, we fitted the fringe profiles with an exponentially-damped sinusoidal function combined with geometrical damped function proportional to $1/\sqrt{L}$.[15,58,69] Accordingly, the polaritonic wavelengths are estimated to be 570 nm for $S_3$ and 660 nm for $\Delta S_2$, corresponding to the ground state and its transient change, respectively.

This result indicates that photocarriers in the thick $WS_2$ layer transiently modulate not only the field amplitude but also the wavelength of HPhPs of hBN. Frequency-dependent measurements (Figure 4e and Supporting Information Figure S6) consistently show longer wavelengths for $\Delta S_2$ compared to $S_3$, confirming the robustness of the red-shift in transient HPhPs.

To theoretically support the experimentally observed ultrafast modulation of HPhPs in hybrid $WS_2$/hBN heterostructures, we performed numerical simulations of both the ground and photoexcited transient states using the finite-element method (FEM), motivated by the procedure described in previous reports.[42,74,75] Figure 5a illustrates the simulation geometries corresponding to the two experimental configurations: $WS_2$ (1.5 nm)/hBN (120 nm)/$SiO_2$ (2000 nm) and hBN (38 nm)/$WS_2$ (376 nm)/$SiO_2$ (1624 nm). The layer thicknesses for these two configurations were taken from Supporting Information Figure S7 and Figure 4a, respectively. We performed the FEM simulation as a two-dimensional simulation. AFM tip is modeled as a dipole acting as the excitation source of phonon polaritons. To simulate the experimentally observed interference pattern of HPhPs, we systematically scanned the dipole position across the $WS_2$/hBN structure from the edge ($X = 0$ μm) to the opposite side. The real part of the electric field along the out-of-plane ($Z$) direction ($\mathrm{Re}[E_Z]$) was evaluated at a point of 100 nm below the dipole. Dielectric functions of hBN, $SiO_2$, and ground-state $WS_2$ were taken from the literature.[76] For photoexcited $WS_2$, we assumed Drude-type response to model the dielectric tensor, such that the in-plane (perpendicular to the $c$-axis: $\varepsilon_\perp$) and out-of-plane (parallel to the $c$-axis: $\varepsilon_\parallel$) components are modified as $\varepsilon_\perp \rightarrow \varepsilon_\perp + \Delta\varepsilon_\perp$, $\varepsilon_\parallel \rightarrow \varepsilon_\parallel + \Delta\varepsilon_\parallel$, with $\Delta\varepsilon_\perp = -5 + 5i$ and $\Delta\varepsilon_\parallel = -0.5 + 0.5i$. $\Delta\varepsilon_\parallel$ values were estimated in our previous study,[67] whereas $\Delta\varepsilon_\perp$ values are assumed empirically. In the Supporting Information Note S4 and Figure S12, we discuss in detail how the different choices of $\Delta\varepsilon_\perp$ and $\Delta\varepsilon_\parallel$ affect the simulated outcomes. For the thick $WS_2$, to account for the finite penetration depth of 515 nm pump beam, we only treated the top 38 nm as an excited domain while leaving the remaining domain to be the ground state.[77]

Figure 5b displays the distribution of $\mathrm{Re}[E_Z]$ in the ground-state of $WS_2$ with the dipole located at $X$ = 1.5 μm (left: $WS_2$/hBN/$SiO_2$ at the frequency of 1497 $cm^{-1}$, right: hBN/$WS_2$/$SiO_2$ at 1425 $cm^{-1}$). Both configurations clearly exhibit oscillatory near-field distributions characteristic of the HPhPs.

We first focus on the $WS_2$/hBN/$SiO_2$ structure as experimentally explored in Figure 3. The top panels of Figure 5c shows the line profiles of $\mathrm{Re}[E_Z]$ under the dipole, calculated for ground-state (GS) and transient-state (TS) of $WS_2$ (denoted as $\mathrm{Re}\left[E_Z^{\mathrm{GS}}\right]$ and $\mathrm{Re}\left[E_Z^{\mathrm{TS}}\right]$, respectively). The bottom panel shows their differential field $\mathrm{Re}[\Delta E_Z] = \mathrm{Re}\left[E_Z^{\mathrm{TS}}\right] - \mathrm{Re}\left[E_Z^{\mathrm{GS}}\right]$ (TS–GS). $\mathrm{Re}\left[E_Z^{\mathrm{TS}}\right]$ exhibits a slightly faster amplitude decay and a marginally longer wavelength ($\lambda_{\mathrm{p}}^{\mathrm{TS}}$ = 1444 nm) compared with $\mathrm{Re}\left[E_Z^{\mathrm{GS}}\right]$ ($\lambda_{\mathrm{p}}^{\mathrm{GS}} = 1440$ nm), resulting in the anti-phase oscillation pattern in $\mathrm{Re}[\Delta E_Z]$. This behavior is clearly consistent with the experimental $\Delta S_2$ profiles shown in Figure 3d.

To evaluate dispersion relationship in the numerical simulations, we calculated $\mathrm{Re}[E_Z]$ at different dipole frequencies, as shown in Supporting Information Figure S8a. From these results, we extracted $\lambda_{\mathrm{p}}$ and plotted the dispersion relation in Supporting Information Figure S8b, where the difference in the imaginary part of $r_{\mathrm{p}}$, defined as $\Delta\mathrm{Im}\left[r_{\mathrm{p}}\right] = \mathrm{Im}\left[r_{\mathrm{p}}^{\mathrm{TS}}\right] - \mathrm{Im}\left[r_{\mathrm{p}}^{\mathrm{GS}}\right]$, is also plotted. The numerically calculated dispersion agrees closely with that reconstructed from ultrafast IR *s*-SNOM measurements.

We note that, in Figure 5c and Supporting Information Figure S8a, the transient modulation of the polaritonic field persists even in the region where $WS_2$ is absent. This behavior arises because the tip-launched polaritonic field propagates through the $WS_2$-covered domain before being reflected at the crystal edge and subsequently detected beneath the dipole. A similar phenomenon is also evident in one of our experimental data sets (Supporting Information Figure S9).

We then investigate the $hBN/WS_2/SiO_2$ configuration as experimentally studied in Figure 4. Figure 5d shows $\mathrm{Re}\left[E_Z^{\mathrm{GS}}\right]$ and $\mathrm{Re}\left[E_Z^{\mathrm{TS}}\right]$ (top), together with $\mathrm{Re}[\Delta E_Z]$ (bottom) for this configuration. The profile of $\mathrm{Re}\left[E_Z^{\mathrm{TS}}\right]$ differs markedly from that of $\mathrm{Re}\left[E_Z^{\mathrm{GS}}\right]$, in contrast to the results for the $WS_2/hBN/SiO_2$ configuration in Figure 5c. This difference arises from the thicker $WS_2$ layer exerting a stronger influence on the HPhPs in the TS state, primarily due to the larger number of photocarriers generated. From fitting analysis of the simulated results, $\lambda_{\mathrm{p}}^{\mathrm{GS}}$ and $\lambda_{\mathrm{p}}^{\mathrm{TS}}$ are determined to be 559 nm and 618 nm, respectively. For $\mathrm{Re}[\Delta E_Z]$, $\lambda_{\mathrm{p}}^{\mathrm{TS-GS}}$ is evaluated to be 599 nm and is longer than $\lambda_{\mathrm{p}}^{\mathrm{GS}}$. This is again consistent with our experimental observation (Figure 4d and 4e) that the wavelength observed in $\Delta S_2$ (corresponding to simulated $\mathrm{Re}[\Delta E_Z]$) is longer than that in $S_3$ (corresponding to simulated $\mathrm{Re}\left[E_Z^{\mathrm{GS}}\right]$).

To further investigate the frequency dependence of the polaritonic wavelength in the $hBN/WS_2/SiO_2$ configuration, we calculated the $\mathrm{Re}[\Delta E_Z]$ at different frequencies and plotted $\lambda_{\mathrm{p}}^{\mathrm{GS}}$ and $\lambda_{\mathrm{p}}^{\mathrm{TS}}$ (Supporting Information Figure S10). The $\lambda_{\mathrm{p}}^{\mathrm{TS-GS}}$ (blue triangles) is consistently longer than that of GS, indicating a redshift of the polariton mode upon photoexcitation.[58] These results show good quantitative agreement with the experimental frequency dependence (Figure 4e).

In Supporting Information, we also examine the pump fluence dependence of the ultrafast modulation of HPhPs in the $hBN/WS_2/SiO_2$ configuration (Supporting Information Note S3 and Figure S11). We experimentally found that the phase of the HPhPs evolved with increasing pump fluence, which we reproduced in our numerical simulation.

We note that a potential complexity arises from the finite propagation time of HPhPs before they are launched by the tip, travel to the crystal edge, get reflected, and detected by the tip. Photocarrier density would temporally decay during this propagation, leading to the enhanced damping of the observed polaritonic profile when the tip is placed away from crystal edges. Supporting Information Figure S13a presents the group velocities ($v_g$) of HPhPs derived from Im $r_p$ in the hBN/$WS_2$/$SiO_2$ structure. The calculated $v_g$ was 1.02 μm/ps at 1425 $cm^{-1}$, corresponding to a propagation time of a few picoseconds within our imaging area (e.g., approximately 4 μm of the distance of tip-edge-tip in Figure 4). Conversely, the carrier lifetime in the sample was determined to be 18 ps based on the fitting of the near-field pump–probe time trace (Supporting Information Figure S13b), which is one order of magnitude larger than propagation time. Consequently, the evolution of the dielectric function of $WS_2$ during the propagation of the HPhPs appears to be relatively minor in our cases. When analyzing the long propagating HPhPs (e.g. edge-excited HPhPs) at a location far from the edge, the effect might become more pronounced.

In summary, our results demonstrate that grating-coupled ultrafast *s*-SNOM visualizes polaritonic dynamics with spatio-temporal-spectral resolution, revealing transient modulation of HPhPs in hBN induced by photocarrier injection into $WS_2$ layers. Simulations confirmed that ultrafast modulation arises from changes in the dielectric function of $WS_2$ under optical excitation, reproducing experimentally observed interference fringes. This approach provides a versatile platform for tailoring and imaging strongly dispersive polaritons in complex vdW stacks at the ultrafast time scale, including actively gated, phase-change, and magnetically tuned systems.

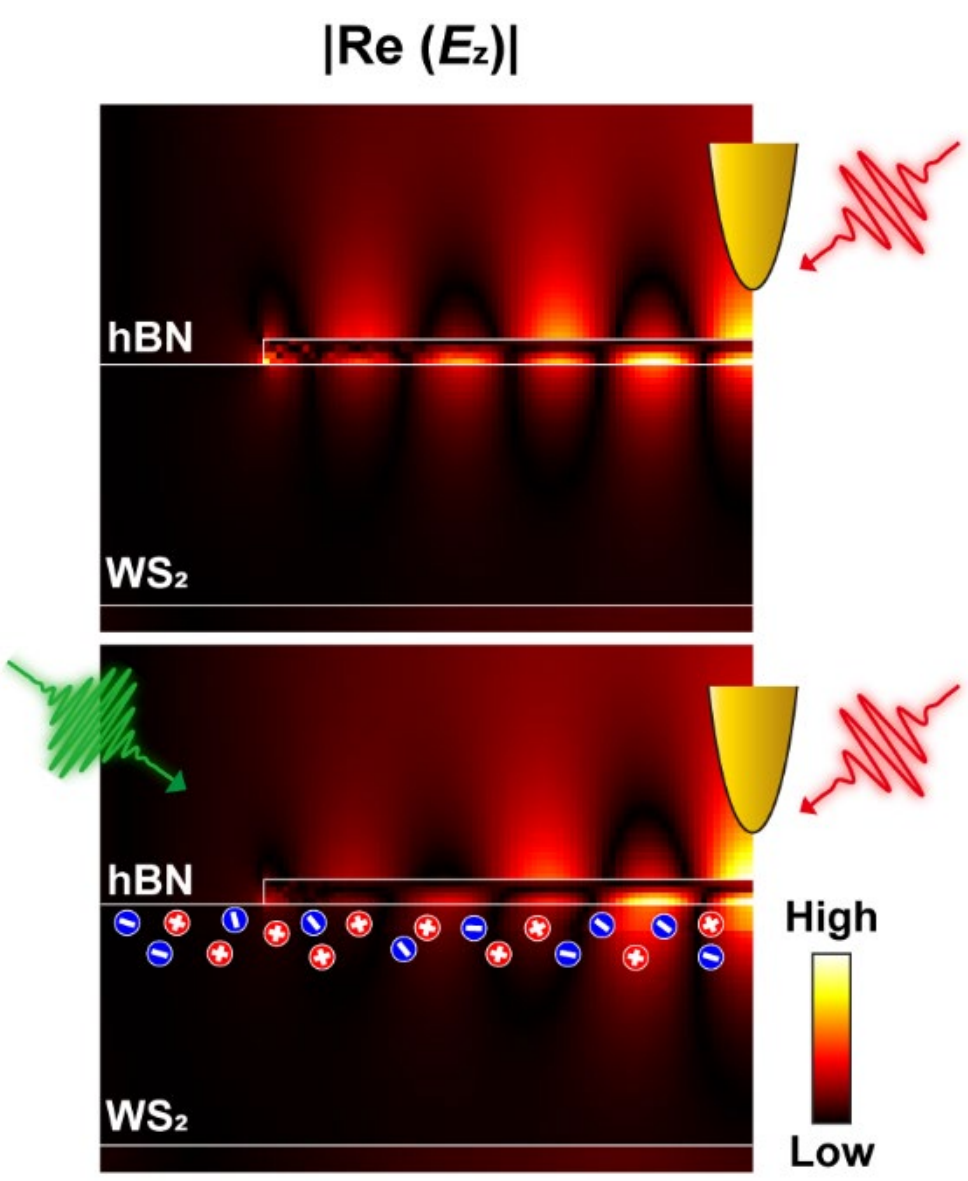

**Figure 1.** Ultrafast control of hyperbolic phonon polaritons. Hyperbolic phonon polaritons in hBN are modulated by visible-pump-induced generation of electron–hole pairs in $WS_2$.

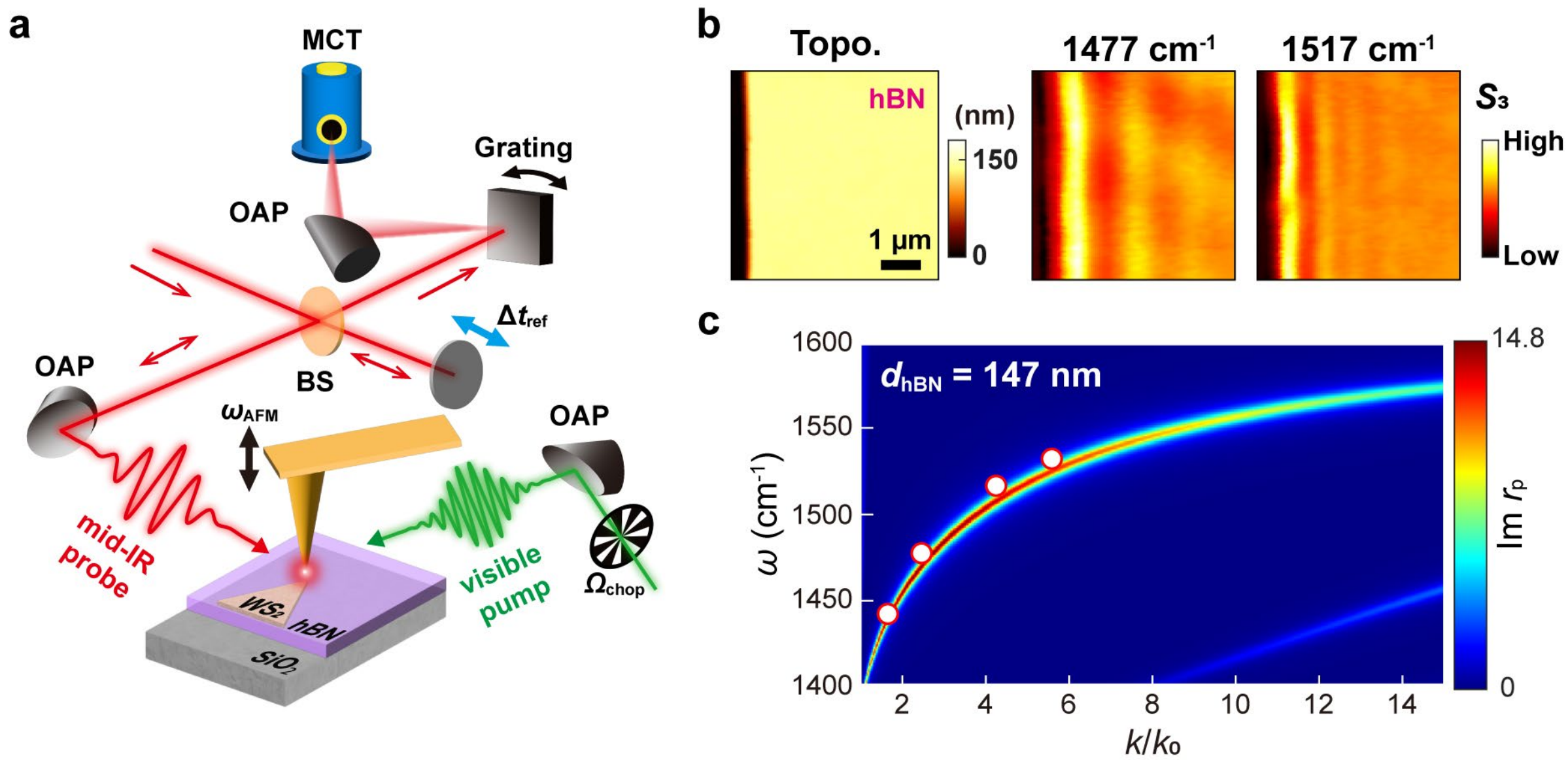

**Figure 2.** Demonstration of polariton imaging using broadband light pulse. (a) Experimental setup for ultrafast nano-imaging of polaritons. BS: ZnSe beam splitter, OAP: off-axis parabolic mirror, $\omega_{AFM}$: tapping frequency of the AFM tip, $\Omega_{chop}$: chopper frequency, $\Delta t_{ref}$: reference delay. (b) AFM topography and SNOM image demodulated at $3\omega_{AFM}$ ($S_3$) of hBN/$SiO_2$ at the frequency of 1477 $cm^{-1}$ and 1517 $cm^{-1}$. (c) Dispersion relationship of phonon polaritons in hBN (147 nm)/$SiO_2$. White circles indicate experimental data obtained by fringe period in (b).

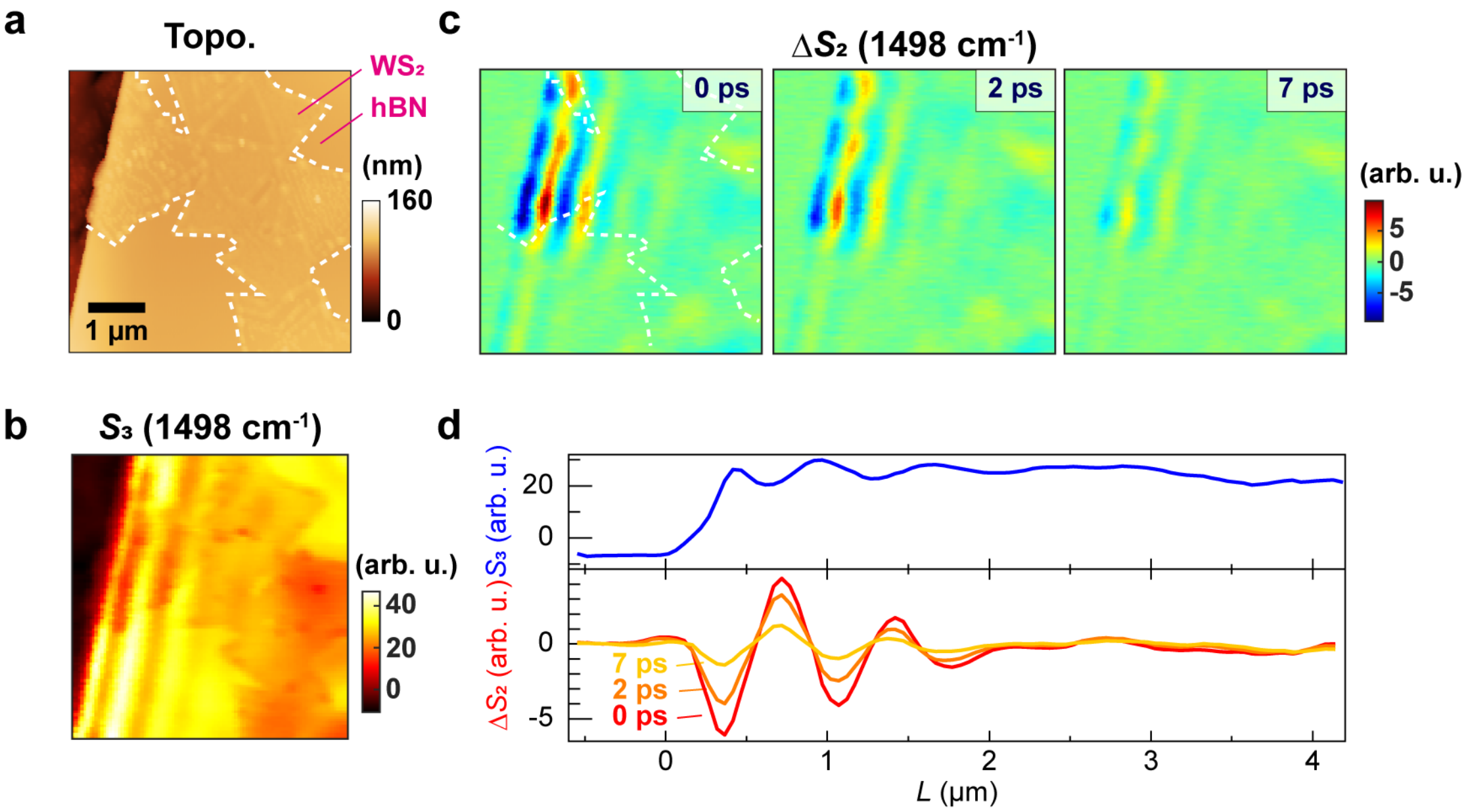

**Figure 3.** Ultrafast pump–probe imaging in $WS_2$/hBN/$SiO_2$ heterostructures. (a) AFM topography of $WS_2$/hBN/$SiO_2$. The domain covered by $WS_2$ is indicated by the white dashed line. (b) A ground-state near-field image at a frequency of 1498 $cm^{-1}$. (c) Transient near-field images at different delay time 0 ps, 2 ps, and 7 ps at a probe frequency of 1498 $cm^{-1}$. (d) Profiles of $S_3$ and time-dependent $\Delta S_2$ extracted from near-field images of (b) and (c).

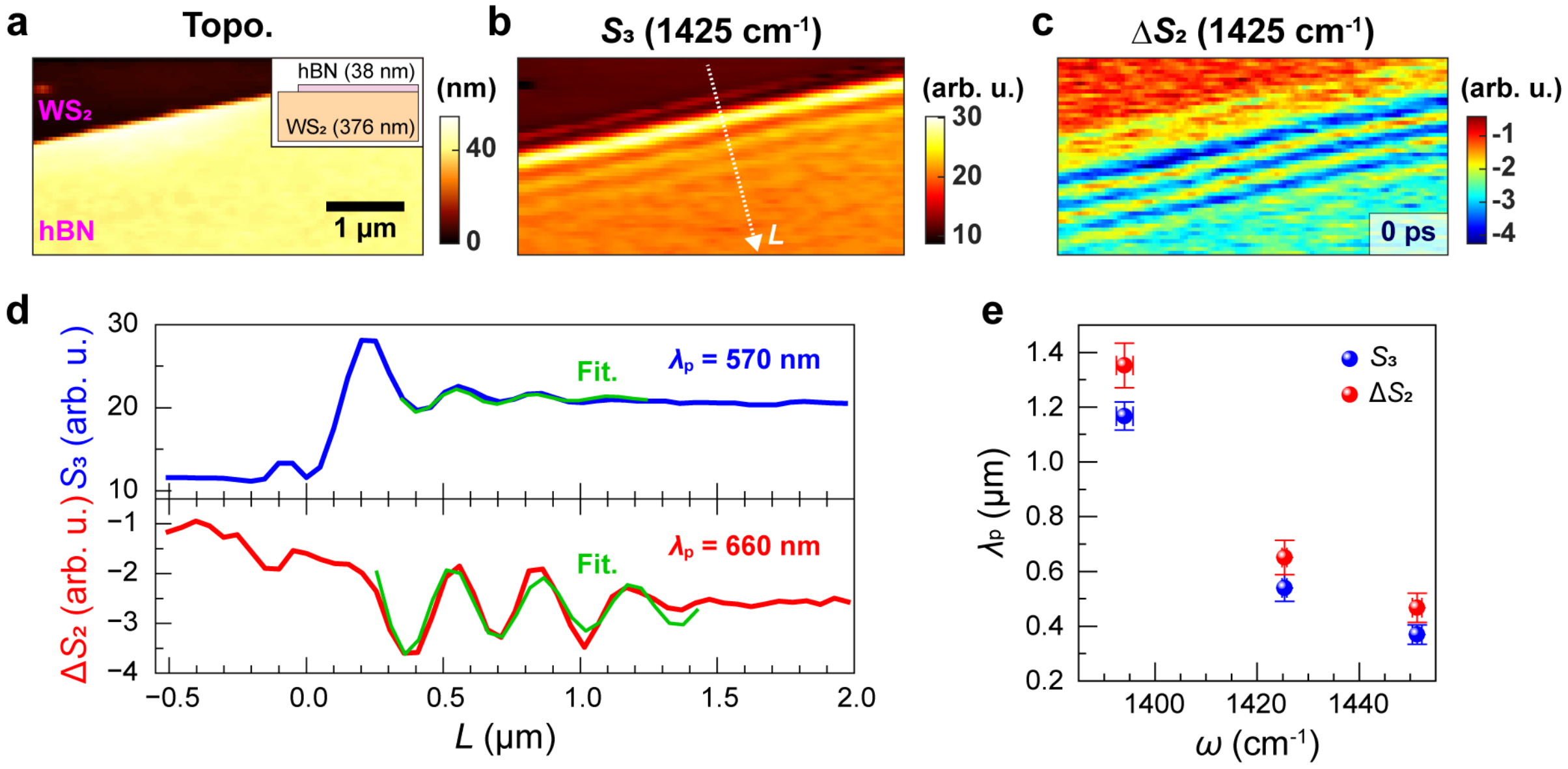

**Figure 4.** Ultrafast pump–probe imaging in $hBN/WS_2/SiO_2$ heterostructures. (a) AFM topography of the $hBN/WS_2/SiO_2$ sample. The thicknesses of the hBN and $WS_2$ layers are 38 nm and 376 nm, respectively. The inset shows the structure of the sample. (b–c) Ground- (b) and transient-state (c) near-field images at a frequency of 1425 $cm^{-1}$. The white dashed line indicates the location where the profile shown in (d) was extracted. (d) Fringe profiles of $S_3$ and $\Delta S_2$ obtained from (b) and (c). The polaritonic wavelengths ($\lambda_p$) were evaluated to be 570 nm and 660 nm for $S_3$ and $\Delta S_2$, respectively, based on the fittings (green line). (e) $\lambda_p$ of $S_3$ and $\Delta S_2$ extracted from the fringe profiles as a function of detection frequency ($\omega$).

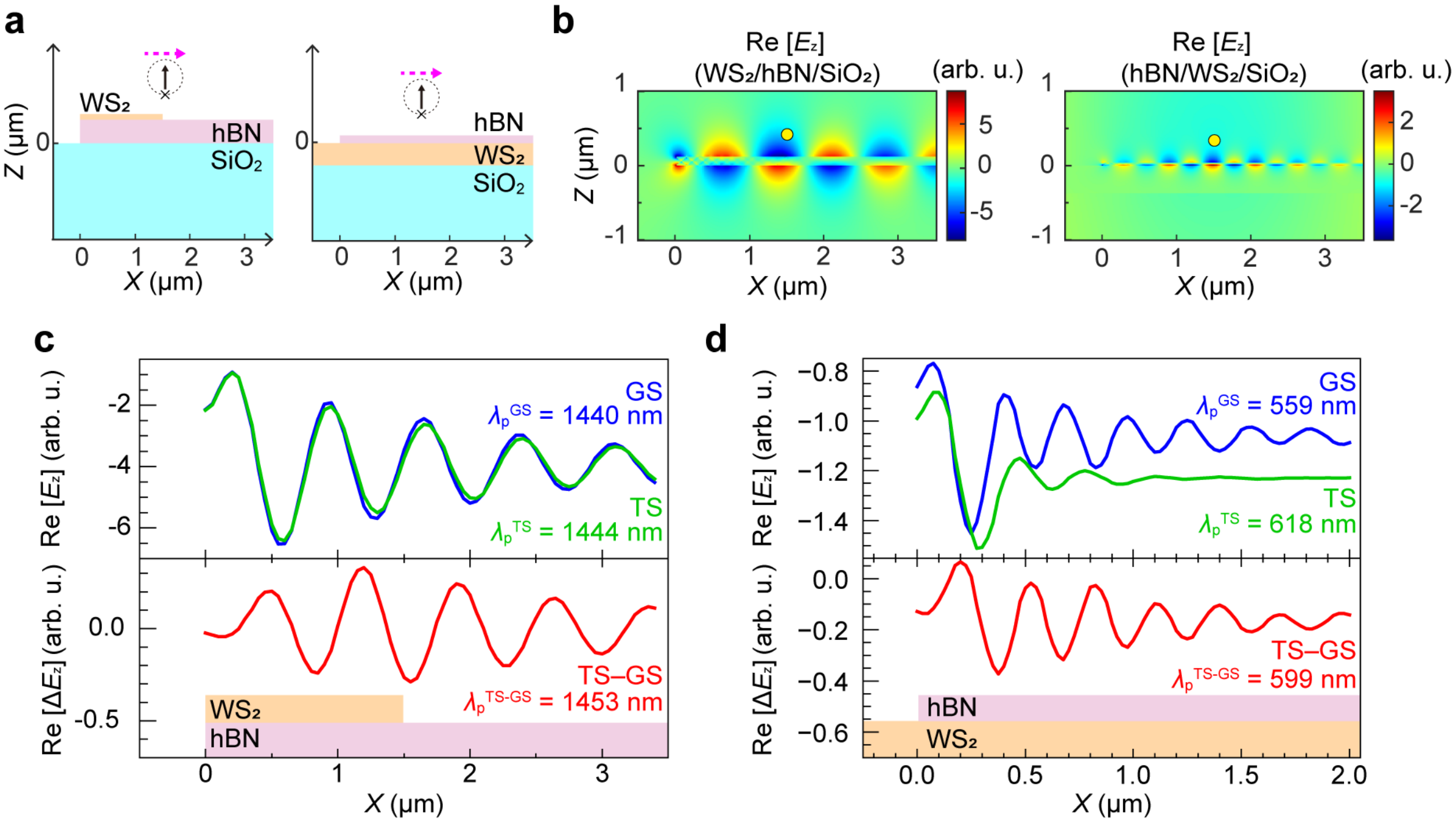

**Figure 5.** Numerical simulations by FEM. (a) Calculation geometry of $WS_2$/hBN/$SiO_2$ (left) and hBN/$WS_2$/$SiO_2$ (right). The dipole source is located at 300 nm above the hBN surface. The field amplitude is monitored at 100 nm below the dipole as shown by a cross mark. (b) Real part of the electric field distribution in *Z* directions, $\mathrm{Re}[E_Z]$ in $WS_2$/hBN/$SiO_2$ (left) and hBN/$WS_2$/$SiO_2$ (right). The dipole location is indicated by yellow circles (c, d) Top: $\mathrm{Re}\left[E_Z^{\mathrm{GS}}\right]$ (blue line) and $\mathrm{Re}\left[E_Z^{\mathrm{TS}}\right]$ (green line), bottom: $\mathrm{Re}[\Delta E_Z]$ (TS–GS) plotted as a function of the dipole-edge distance *X*. The schematics of (c) $WS_2$/hBN and (d) hBN/$WS_2$ used in the calculation are shown at the bottom of each figure.

ASSOCIATED CONTENT

**Supporting Information**.

Experimental details. A near-field image in broadband detection. Observation of edge-launched polaritons. Near-field pump–probe time trace in narrowband detection. Frequency dependence of near-field pump–probe images. Another data set of near-field images. Dipole frequency dependence of simulation results. Pump-fluence dependence. The effect of dielectric function of $WS_2$ on simulation results. Group velocities of HPhPs and near-field time trace. (PDF)

AUTHOR INFORMATION

**Corresponding Author**

*Takashi Kumagai

e-mail: kuma@ims.ac.jp

ORCID: 0000-0001-7029-062X

*Jun Nishida

e-mail: nishida@ims.ac.jp

ORCID: 0000-0001-7834-8179

**Author Contributions**

J.N. conceived the project. K.S., F.Z., K.W., T.T., and R.K. prepared the samples. K.K., T.K., and J.N. performed the spectroscopy measurements. K.K., A.P., T.K., and J.N. analyzed the data.

M.S., T.K., and J.N. supervised the project. K.K., T.K., and J.N. wrote the original draft. K.K, K.S., R.K., A.P., M.S., T.K., and J.N. reviewed and edited the manuscript.

**Notes**

The authors declare no competing financial interest.

ACKNOWLEDGMENT

This work was supported by JSPS KAKENHI (JP24K01443, JP22K14653, and JP25H01395 to J.N.; JP24H01209 to T.K.; JP25K00936 to K.S.; JP23H05469 to R.K.; and JP21H05233 and JP23H02052 to K.W. and T.T.); the Japan Science and Technology Agency FOREST Program (JPMJFR236L to J.N., JPMJFR201J to T.K., and JPMJFR213K to K.S.); JST CREST (JPMJCR24A5 to K.W. and T.T.); the World Premier International Research Center Initiative (WPI), MEXT, Japan (K.W. and T.T.); the Research Foundation for Opto-Science and Technology (J.N.); the Frontier Photonic Sciences Project of the National Institutes of Natural Sciences (Grant 01213008 to J.N.); and the Daiko Foundation (J.N.).

REFERENCES

(1) Caldwell, J. D.; Lindsay, L.; Giannini, V.; Vurgaftman, I.; Reinecke, T. L.; Maier, S. A.; Glembocki, O. J. Low-Loss, Infrared and Terahertz Nanophotonics Using Surface Phonon Polaritons. *Nanophotonics* **2015**, *4* (1), 44–68.

(2) Dai, S.; Fei, Z.; Ma, Q.; Rodin, A. S.; Wagner, M.; McLeod, A. S.; Liu, M. K.; Gannett, W.; Regan, W.; Watanabe, K.; Taniguchi, T.; Thiemens, M.; Dominguez, G.; Neto, A. H. C.; Zettl, A.; Keilmann, F.; Jarillo-Herrero, P.; Fogler, M. M.; Basov, D. N. Tunable Phonon Polaritons in Atomically Thin van der Waals Crystals of Boron Nitride. *Science* **2014**, *343* (6175), 1125–1129.

(3) Caldwell, J. D.; Kretinin, A. V.; Chen, Y.; Giannini, V.; Fogler, M. M.; Francescato, Y.; Ellis, C. T.; Tischler, J. G.; Woods, C. R.; Giles, A. J.; Hong, M.; Watanabe, K.; Taniguchi, T.; Maier, S. A.; Novoselov, K. S. Sub-Diffractional Volume-Confined Polaritons in the Natural Hyperbolic Material Hexagonal Boron Nitride. *Nat. Commun.* **2014**, *5* (1), 5221.

(4) Li, P.; Lewin, M.; Kretinin, A. V.; Caldwell, J. D.; Novoselov, K. S.; Taniguchi, T.; Watanabe, K.; Gaussmann, F.; Taubner, T. Hyperbolic Phonon-Polaritons in Boron Nitride for Near-Field Optical Imaging and Focusing. *Nat. Commun.* **2015**, *6* (1), 7507.

(5) Yoxall, E.; Schnell, M.; Nikitin, A. Y.; Txoperena, O.; Woessner, A.; Lundeberg, M. B.; Casanova, F.; Hueso, L. E.; Koppens, F. H. L.; Hillenbrand, R. Direct Observation of Ultraslow Hyperbolic Polariton Propagation with Negative Phase Velocity. *Nat. Photonics* **2015**, *9* (10), 674–678.

(6) Autore, M.; Li, P.; Dolado, I.; Alfaro-Mozaz, F. J.; Esteban, R.; Atxabal, A.; Casanova, F.; Hueso, L. E.; Alonso-González, P.; Aizpurua, J.; Nikitin, A. Y.; Vélez, S.; Hillenbrand, R. Boron

Nitride Nanoresonators for Phonon-Enhanced Molecular Vibrational Spectroscopy at the Strong Coupling Limit. *Light Sci. Appl.* **2018**, *7* (4), 17172–17172.

(7) Hutchins, W.; Zare, S.; Hirt, D. M.; Tomko, J. A.; Matson, J. R.; Diaz-Granados, K.; Long, M.; He, M.; Pfeifer, T.; Li, J.; Edgar, J. H.; Maria, J.-P.; Caldwell, J. D.; Hopkins, P. E. Ultrafast Evanescent Heat Transfer across Solid Interfaces via Hyperbolic Phonon–Polariton Modes in Hexagonal Boron Nitride. *Nat. Mater.* **2025**, *24* (5), 698–706.

(8) Poddubny, A.; Iorsh, I.; Belov, P.; Kivshar, Y. Hyperbolic Metamaterials. *Nat. Photonics* **2013**, *7* (12), 948–957.

(9) Giles, A. J.; Dai, S.; Vurgaftman, I.; Hoffman, T.; Liu, S.; Lindsay, L.; Ellis, C. T.; Assefa, N.; Chatzakis, I.; Reinecke, T. L.; Tischler, J. G.; Fogler, M. M.; Edgar, J. H.; Basov, D. N.; Caldwell, J. D. Ultralow-Loss Polaritons in Isotopically Pure Boron Nitride. *Nature Mater.* **2018**, *17* (2), 134–139.

(10) Giles, A. J.; Dai, S.; Glembocki, O. J.; Kretinin, A. V.; Sun, Z.; Ellis, C. T.; Tischler, J. G.; Taniguchi, T.; Watanabe, K.; Fogler, M. M.; Novoselov, K. S.; Basov, Dimitri. N.; Caldwell, J. D. Imaging of Anomalous Internal Reflections of Hyperbolic Phonon-Polaritons in Hexagonal Boron Nitride. *Nano Lett.* **2016**, *16* (6), 3858–3865.

(11) Li, P.; Dolado, I.; Alfaro-Mozaz, F. J.; Nikitin, A. Yu.; Casanova, F.; Hueso, L. E.; Vélez, S.; Hillenbrand, R. Optical Nanoimaging of Hyperbolic Surface Polaritons at the Edges of van der Waals Materials. *Nano Lett.* **2017**, *17* (1), 228–235.

(12) Menabde, S. G.; Boroviks, S.; Ahn, J.; Heiden, J. T.; Watanabe, K.; Taniguchi, T.; Low, T.; Hwang, D. K.; Mortensen, N. A.; Jang, M. S. Near-Field Probing of Image Phonon-Polaritons in Hexagonal Boron Nitride on Gold Crystals. *Science Advances* **2022**, *8* (28), eabn0627.

(13) Ma, W.; Alonso-González, P.; Li, S.; Nikitin, A. Y.; Yuan, J.; Martín-Sánchez, J.; Taboada-Gutiérrez, J.; Amenabar, I.; Li, P.; Vélez, S.; Tollan, C.; Dai, Z.; Zhang, Y.; Sriram, S.; Kalantar-Zadeh, K.; Lee, S.-T.; Hillenbrand, R.; Bao, Q. In-Plane Anisotropic and Ultra-Low-Loss Polaritons in a Natural van der Waals Crystal. *Nature* **2018**, *562* (7728), 557–562.

(14) Zheng, Z.; Chen, J.; Wang, Y.; Wang, X.; Chen, X.; Liu, P.; Xu, J.; Xie, W.; Chen, H.; Deng, S.; Xu, N. Highly Confined and Tunable Hyperbolic Phonon Polaritons in van der Waals Semiconducting Transition Metal Oxides. *Adv. Mater.* **2018**, *30* (13), 1705318.

(15) Taboada-Gutiérrez, J.; Álvarez-Pérez, G.; Duan, J.; Ma, W.; Crowley, K.; Prieto, I.; Bylinkin, A.; Autore, M.; Volkova, H.; Kimura, K.; Kimura, T.; Berger, M.-H.; Li, S.; Bao, Q.; Gao, X. P. A.; Errea, I.; Nikitin, A. Y.; Hillenbrand, R.; Martín-Sánchez, J.; Alonso-González, P. Broad Spectral Tuning of Ultra-Low-Loss Polaritons in a van der Waals Crystal by Intercalation. *Nat. Mater.* **2020**, *19* (9), 964–968.

(16) Passler, N. C.; Ni, X.; Hu, G.; Matson, J. R.; Carini, G.; Wolf, M.; Schubert, M.; Alù, A.; Caldwell, J. D.; Folland, T. G.; Paarmann, A. Hyperbolic Shear Polaritons in Low-Symmetry Crystals. *Nature* **2022**, *602* (7898), 595–600.

(17) Matson, J.; Wasserroth, S.; Ni, X.; Obst, M.; Diaz-Granados, K.; Carini, G.; Renzi, E. M.; Galiffi, E.; Folland, T. G.; Eng, L. M.; Michael Klopf, J.; Mastel, S.; Armster, S.; Gambin, V.; Wolf, M.; Kehr, S. C.; Alù, A.; Paarmann, A.; Caldwell, J. D. Controlling the Propagation

Asymmetry of Hyperbolic Shear Polaritons in Beta-Gallium Oxide. *Nat. Commun.* **2023**, *14* (1), 5240.

(18) Hu, G.; Ma, W.; Hu, D.; Wu, J.; Zheng, C.; Liu, K.; Zhang, X.; Ni, X.; Chen, J.; Zhang, X.; Dai, Q.; Caldwell, J. D.; Paarmann, A.; Alù, A.; Li, P.; Qiu, C.-W. Real-Space Nanoimaging of Hyperbolic Shear Polaritons in a Monoclinic Crystal. *Nat. Nanotechnol.* **2023**, *18* (1), 64–70.

(19) Díaz-Núñez, P.; Lanza, C.; Wang, Z.; Kravets, V. G.; Duan, J.; Álvarez-Cuervo, J.; Martín-Luengo, A. T.; Grigorenko, A. N.; Yang, Q.; Paarmann, A.; Caldwell, J.; Alonso-González, P.; Mishchenko, A. Visualization of Topological Shear Polaritons in Gypsum Thin Films. *Science Advances* **2025**, *11* (29), eadw3452.

(20) Ma, W.; Hu, G.; Hu, D.; Chen, R.; Sun, T.; Zhang, X.; Dai, Q.; Zeng, Y.; Alù, A.; Qiu, C.-W.; Li, P. Ghost Hyperbolic Surface Polaritons in Bulk Anisotropic Crystals. *Nature* **2021**, *596* (7872), 362–366.

(21) Suriyage, M.; Zhou, Q.; Qin, H.; Sun, X.; Lu, Z.; Maier, S. A.; Yu, Z.; Lu, Y. Long-Propagating Ghost Phonon Polaritons Enabled by Selective Mode Excitation. *Light Sci. Appl.* **2025**, *14* (1), 254.

(22) Feres, F. H.; Mayer, R. A.; Wehmeier, L.; Maia, F. C. B.; Viana, E. R.; Malachias, A.; Bechtel, H. A.; Klopf, J. M.; Eng, L. M.; Kehr, S. C.; González, J. C.; Freitas, R. O.; Barcelos, I. D. Sub-Diffractional Cavity Modes of Terahertz Hyperbolic Phonon Polaritons in Tin Oxide. *Nat. Commun.* **2021**, *12* (1), 1995.

(23) F. Tresguerres-Mata, A. I.; Lanza, C.; Taboada-Gutiérrez, J.; Matson, Joseph. R.; Álvarez-Pérez, G.; Isobe, M.; Tarazaga Martín-Luengo, A.; Duan, J.; Partel, S.; Vélez, M.; Martín-Sánchez,

J.; Nikitin, A. Y.; Caldwell, J. D.; Alonso-González, P. Observation of Naturally Canalized Phonon Polaritons in $LiV_2O_5$ Thin Layers. *Nat. Commun.* **2024**, *15* (1), 2696.

(24) Zheng, C.; Hu, G.; Wei, J.; Ma, X.; Li, Z.; Chen, Y.; Ni, Z.; Li, P.; Wang, Q.; Qiu, C.-W. Hyperbolic-to-Hyperbolic Transition at Exceptional Reststrahlen Point in Rare-Earth Oxyorthosilicates. *Nat. Commun.* **2024**, *15* (1), 7047.

(25) Sun, T.; Chen, R.; Ma, W.; Wang, H.; Yan, Q.; Luo, J.; Zhao, S.; Zhang, X.; Li, P. van der Waals Quaternary Oxides for Tunable Low-Loss Anisotropic Polaritonics. *Nat. Nanotechnol.* **2024**, *19* (6), 758–765.

(26) Kowalski, R. A.; Mueller, N. S.; Álvarez-Pérez, G.; Obst, M.; Diaz-Granados, K.; Carini, G.; Senarath, A.; Dixit, S.; Niemann, R.; Iyer, R. B.; Kaps, F. G.; Wetzel, J.; Klopf, J. M.; Kravchenko, I. I.; Wolf, M.; Folland, T. G.; Eng, L. M.; Kehr, S. C.; Alonso-Gonzalez, P.; Paarmann, A.; Caldwell, J. D. Ultraconfined Terahertz Phonon Polaritons in Hafnium Dichalcogenides. *Nat. Mater.* **2025**, *24* (11), 1735–1741.

(27) Galiffi, E.; Carini, G.; Ni, X.; Álvarez-Pérez, G.; Yves, S.; Renzi, E. M.; Nolen, R.; Wasserroth, S.; Wolf, M.; Alonso-Gonzalez, P.; Paarmann, A.; Alù, A. Extreme Light Confinement and Control in Low-Symmetry Phonon-Polaritonic Crystals. *Nat. Rev. Mater.* **2024**, *9* (1), 9–28.

(28) Hu, G.; Shen, J.; Qiu, C.-W.; Alù, A.; Dai, S. Phonon Polaritons and Hyperbolic Response in van der Waals Materials. *Advanced Optical Materials* **2020**, *8* (5), 1901393.

(29) Hillenbrand, R.; Abate, Y.; Liu, M.; Chen, X.; Basov, D. N. Visible-to-THz Near-Field Nanoscopy. *Nat. Rev. Mater.* **2025**, *10* (4), 285–310.

(30) Basov, D. N.; Fogler, M. M.; García De Abajo, F. J. Polaritons in van der Waals Materials. *Science* **2016**, *354* (6309), aag1992.

(31) Kim, K. S.; Trajanoski, D.; Ho, K.; Gilburd, L.; Maiti, A.; van der Velden, L.; de Beer, S.; Walker, G. C. The Effect of Adjacent Materials on the Propagation of Phonon Polaritons in Hexagonal Boron Nitride. *J. Phys. Chem. Lett.* **2017**, *8* (13), 2902–2908.

(32) Fali, A.; White, S. T.; Folland, T. G.; He, M.; Aghamiri, N. A.; Liu, S.; Edgar, J. H.; Caldwell, J. D.; Haglund, R. F.; Abate, Y. Refractive Index-Based Control of Hyperbolic Phonon-Polariton Propagation. *Nano Lett.* **2019**, *19* (11), 7725–7734.

(33) Feres, F. H.; Mayer, R. A.; Barcelos, I. D.; Freitas, R. O.; Maia, F. C. B. Acceleration of Subwavelength Polaritons by Engineering Dielectric-Metallic Substrates. *ACS Photonics* **2020**, *7* (6), 1396–1402.

(34) Lee, I.-H.; He, M.; Zhang, X.; Luo, Y.; Liu, S.; Edgar, J. H.; Wang, K.; Avouris, P.; Low, T.; Caldwell, J. D.; Oh, S.-H. Image Polaritons in Boron Nitride for Extreme Polariton Confinement with Low Losses. *Nat. Commun.* **2020**, *11* (1), 3649.

(35) Duan, J.; Álvarez-Pérez, G.; Tresguerres-Mata, A. I. F.; Taboada-Gutiérrez, J.; Voronin, K. V.; Bylinkin, A.; Chang, B.; Xiao, S.; Liu, S.; Edgar, J. H.; Martín, J. I.; Volkov, V. S.; Hillenbrand, R.; Martín-Sánchez, J.; Nikitin, A. Y.; Alonso-González, P. Planar Refraction and Lensing of Highly Confined Polaritons in Anisotropic Media. *Nat. Commun.* **2021**, *12* (1), 4325.

(36) He, M.; Halimi, S. I.; Folland, T. G.; Sunku, S. S.; Liu, S.; Edgar, J. H.; Basov, D. N.; Weiss, S. M.; Caldwell, J. D. Guided Mid-IR and Near-IR Light within a Hybrid Hyperbolic-Material/Silicon Waveguide Heterostructure. *Advanced Materials* **2021**, *33* (11), 2004305.

(37) Dai, S.; Quan, J.; Hu, G.; Qiu, C.-W.; Tao, T. H.; Li, X.; Alù, A. Hyperbolic Phonon Polaritons in Suspended Hexagonal Boron Nitride. *Nano Lett.* **2019**, *19* (2), 1009–1014.

(38) Shen, J.; Zheng, Z.; Dinh, T.; Wang, C.; Chen, M.; Chen, P.; Ma, Q.; Jarillo-Herrero, P.; Kang, L.; Dai, S. Hyperbolic Phonon Polaritons with Positive and Negative Phase Velocities in Suspended α-$MoO_3$. *Appl. Phys. Lett.* **2022**, *120* (11), 113101.

(39) Hu, G.; Ou, Q.; Si, G.; Wu, Y.; Wu, J.; Dai, Z.; Krasnok, A.; Mazor, Y.; Zhang, Q.; Bao, Q.; Qiu, C.-W.; Alù, A. Topological Polaritons and Photonic Magic Angles in Twisted α-$MoO_3$ Bilayers. *Nature* **2020**, *582* (7811), 209–213.

(40) Chaudhary, K.; Tamagnone, M.; Rezaee, M.; Bediako, D. K.; Ambrosio, A.; Kim, P.; Capasso, F. Engineering Phonon Polaritons in van der Waals Heterostructures to Enhance In-Plane Optical Anisotropy. *Science Advances* **2019**, *5* (4), eaau7171.

(41) Zhang, S.; Ma, P.; You, O.; Zhou, S.; Feng, K.; Yuan, H.; Zhang, J.; Wu, C.; Luo, Y.; Yang, B.; Qiu, C.-W.; Yang, X.; Guo, X.; Liu, Y.; Zhang, S.; Dai, Q. Phonon Engineering Enables Hyperbolic Asymptotic Line Polaritons. *Nat. Nanotechnol.* **2025**, *21*, 223–228.

(42) Bylinkin, A.; Schnell, M.; Autore, M.; Calavalle, F.; Li, P.; Taboada-Gutièrrez, J.; Liu, S.; Edgar, J. H.; Casanova, F.; Hueso, L. E.; Alonso-Gonzalez, P.; Nikitin, A. Y.; Hillenbrand, R. Real-Space Observation of Vibrational Strong Coupling between Propagating Phonon Polaritons and Organic Molecules. *Nat. Photonics* **2021**, *15* (3), 197–202.

(43) F. Tresguerres-Mata, A. I.; Matveeva, O. G.; Lanza, C.; Álvarez-Cuervo, J.; Voronin, K. V.; Calavalle, F.; Avedissian, G.; Díaz-Núñez, P.; Álvarez-Pérez, G.; Tarazaga Martín-Luengo, A.; Taboada-Gutiérrez, J.; Duan, J.; Martín-Sánchez, J.; Bylinkin, A.; Hillenbrand, R.;

Mishchenko, A.; Hueso, L. E.; Volkov, V. S.; Nikitin, A. Y.; Alonso-González, P. Directional Strong Coupling at the Nanoscale between Hyperbolic Polaritons and Organic Molecules. *Nat. Photonics* **2025**, *19*, 1196–1202.

(44) Folland, T. G.; Fali, A.; White, S. T.; Matson, J. R.; Liu, S.; Aghamiri, N. A.; Edgar, J. H.; Haglund, R. F.; Abate, Y.; Caldwell, J. D. Reconfigurable Infrared Hyperbolic Metasurfaces Using Phase Change Materials. *Nat. Commun.* **2018**, *9* (1), 4371.

(45) Dai, S.; Zhang, J.; Ma, Q.; Kittiwatanakul, S.; McLeod, A.; Chen, X.; Corder, S. G.; Watanabe, K.; Taniguchi, T.; Lu, J.; Dai, Q.; Jarillo-Herrero, P.; Liu, M.; Basov, D. N. Phase-Change Hyperbolic Heterostructures for Nanopolaritonics: A Case Study of hBN/$VO_2$. *Advanced Materials* **2019**, *31* (18), 1900251.

(46) Jäckering, L.; Moos, A.; Conrads, L.; Li, Y.; Rothstein, A.; Malik, D.; Watanabe, K.; Taniguchi, T.; Wuttig, M.; Stampfer, C.; Taubner, T. Tailoring Phonon Polaritons in hBN with the Plasmonic Phase-Change Material $In_3SbTe_2$. *Nano Lett.* **2025**, *25* (44), 15809–15816.

(47) Li, P.; Yang, X.; Maß, T. W. W.; Hanss, J.; Lewin, M.; Michel, A.-K. U.; Wuttig, M.; Taubner, T. Reversible Optical Switching of Highly Confined Phonon–Polaritons with an Ultrathin Phase-Change Material. *Nature Mater.* **2016**, *15* (8), 870–875.

(48) Aghamiri, N. A.; Hu, G.; Fali, A.; Zhang, Z.; Li, J.; Balendhran, S.; Walia, S.; Sriram, S.; Edgar, J. H.; Ramanathan, S.; Alù, A.; Abate, Y. Reconfigurable Hyperbolic Polaritonics with Correlated Oxide Metasurfaces. *Nat. Commun.* **2022**, *13* (1), 4511.

(49) Dai, S.; Ma, Q.; Liu, M. K.; Andersen, T.; Fei, Z.; Goldflam, M. D.; Wagner, M.; Watanabe, K.; Taniguchi, T.; Thiemens, M.; Keilmann, F.; Janssen, G. C. a. M.; Zhu, S.-E.; Jarillo-

Herrero, P.; Fogler, M. M.; Basov, D. N. Graphene on Hexagonal Boron Nitride as a Tunable Hyperbolic Metamaterial. *Nat. Nanotechnol.* **2015**, *10* (8), 682–686.

(50) Álvarez-Pérez, G.; González-Morán, A.; Capote-Robayna, N.; Voronin, K. V.; Duan, J.; Volkov, V. S.; Alonso-González, P.; Nikitin, A. Y. Active Tuning of Highly Anisotropic Phonon Polaritons in van der Waals Crystal Slabs by Gated Graphene. *ACS Photonics* **2022**, *9* (2), 383–390.

(51) Zeng, Y.; Ou, Q.; Liu, L.; Zheng, C.; Wang, Z.; Gong, Y.; Liang, X.; Zhang, Y.; Hu, G.; Yang, Z.; Qiu, C.-W.; Bao, Q.; Chen, H.; Dai, Z. Tailoring Topological Transitions of Anisotropic Polaritons by Interface Engineering in Biaxial Crystals. *Nano Lett.* **2022**, *22* (10), 4260–4268.

(52) Hu, H.; Chen, N.; Teng, H.; Yu, R.; Xue, M.; Chen, K.; Xiao, Y.; Qu, Y.; Hu, D.; Chen, J.; Sun, Z.; Li, P.; de Abajo, F. J. G.; Dai, Q. Gate-Tunable Negative Refraction of Mid-Infrared Polaritons. *Science* **2023**, *379* (6632), 558–561.

(53) Mayer, R. A.; Chen, X.; Jing, R.; Tsuneto, M.; Zhou, B.; Zhou, Z.; Zheng, W.; Pu, R.; Xu, S.; Liu, T.; Yao, H.; Wehmeier, L.; Dong, Y.; Sun, D.; He, L.; Cadore, A. R.; Heinz, T.; Fan, J. A.; Dean, C. R.; Basov, D. N.; Du, X.; Freitas, R. O.; Liu, M. Magnetically Tunable Polariton Cavities in van der Waals Heterostructures. *Nano Lett.* **2025**, *25* (35), 13079–13086.

(54) Zhang, Q.; Hu, G.; Ma, W.; Li, P.; Krasnok, A.; Hillenbrand, R.; Alù, A.; Qiu, C.-W. Interface Nano-Optics with van der Waals Polaritons. *Nature* **2021**, *597* (7875), 187–195.

(55) Sternbach, A. J.; Chae, S. H.; Latini, S.; Rikhter, A. A.; Shao, Y.; Li, B.; Rhodes, D.; Kim, B.; Schuck, P. J.; Xu, X. Programmable Hyperbolic Polaritons in van der Waals Semiconductors. *Science* **2021**, 371 (6529), 617–620.

(56) Fu, R.; Qu, Y.; Xue, M.; Liu, X.; Chen, S.; Zhao, Y.; Chen, R.; Li, B.; Weng, H.; Liu, Q.; Dai, Q.; Chen, J. Manipulating Hyperbolic Transient Plasmons in a Layered Semiconductor. *Nat. Commun.* **2024**, *15* (1), 709.

(57) Zhang, X.; Yan, Q.; Ma, W.; Zhang, T.; Yang, X.; Zhang, X.; Li, P. Ultrafast Anisotropic Dynamics of Hyperbolic Nanolight Pulse Propagation. *Science Advances* **2023**, *9* (34), eadi4407.

(58) He, M.; Matson, J. R.; Yu, M.; Cleri, A.; Sunku, S. S.; Janzen, E.; Mastel, S.; Folland, T. G.; Edgar, J. H.; Basov, D. N.; Maria, J.-P.; Law, S.; Caldwell, J. D. Polariton Design and Modulation via van der Waals/Doped Semiconductor Heterostructures. *Nat. Commun.* **2023**, *14* (1), 7965.

(59) Ni, G. X.; Wang, L.; Goldflam, M. D.; Wagner, M.; Fei, Z.; McLeod, A. S.; Liu, M. K.; Keilmann, F.; Özyilmaz, B.; Castro Neto, A. H.; Hone, J.; Fogler, M. M.; Basov, D. N. Ultrafast Optical Switching of Infrared Plasmon Polaritons in High-Mobility Graphene. *Nat. Photonics* **2016**, *10* (4), 244–247.

(60) Vicentini, E.; Arrieta, X.; Schnell, M.; Pajusco, N.; Begemann, F.; Burillo, M. B.; Ramos, M.; Bylinkin, A.; Esteban, R.; Aizpurua, J.; Hillenbrand, R. Real-Space Observation of Flat-Band Ultrastrong Coupling between Optical Phonons and Surface Plasmon Polaritons. *Nat. Mater.* **2026**, *25*, 216–222.

(61) Huber, M. A.; Mooshammer, F.; Plankl, M.; Viti, L.; Sandner, F.; Kastner, L. Z.; Frank, T.; Fabian, J.; Vitiello, M. S.; Cocker, T. L.; Huber, R. Femtosecond Photo-Switching of Interface Polaritons in Black Phosphorus Heterostructures. *Nat. Nanotechnol.* **2017**, *12* (3), 207–211.

(62) Fei, Z.; Rodin, A. S.; Andreev, G. O.; Bao, W.; McLeod, A. S.; Wagner, M.; Zhang, L. M.; Zhao, Z.; Thiemens, M.; Dominguez, G.; Fogler, M. M.; Neto, A. H. C.; Lau, C. N.; Keilmann, F.; Basov, D. N. Gate-Tuning of Graphene Plasmons Revealed by Infrared Nano-Imaging. *Nature* **2012**, *487* (7405), 82–85.

(63) Chen, J.; Badioli, M.; Alonso-González, P.; Thongrattanasiri, S.; Huth, F.; Osmond, J.; Spasenović, M.; Centeno, A.; Pesquera, A.; Godignon, P.; Zurutuza Elorza, A.; Camara, N.; de Abajo, F. J. G.; Hillenbrand, R.; Koppens, F. H. L. Optical Nano-Imaging of Gate-Tunable Graphene Plasmons. *Nature* **2012**, *487* (7405), 77–81.

(64) Mrejen, M.; Yadgarov, L.; Levanon, A.; Suchowski, H. Transient Exciton-Polariton Dynamics in $WSe_2$ by Ultrafast Near-field Imaging. *Science Advances* **2019**, *5* (2), eaat9618.

(65) Nishida, J.; Johnson, S. C.; Chang, P. T. S.; Wharton, D. M.; Dönges, S. A.; Khatib, O.; Raschke, M. B. Ultrafast Infrared Nano-Imaging of Far-from-Equilibrium Carrier and Vibrational Dynamics. *Nat. Commun.* **2022**, *13* (1), 1083.

(66) Nishida, J.; Otsuka, K.; Minato, T.; Kato, Y. K.; Kumagai, T. Ultrafast Infrared Nano-Imaging of Local Electron-Hole Dynamics in CVD-Grown Single-Walled Carbon Nanotubes. *Science Advances* **2025**, *11* (25), eadv9584.

(67) Wang, Y.; Nishida, J.; Nakamoto, K.; Yang, X.; Sakuma, Y.; Zhang, W.; Endo, T.; Miyata, Y.; Kumagai, T. Ultrafast Nano-Imaging of Spatially Modulated Many-Body Dynamics in CVD-Grown Monolayer $WS_2$. *ACS Photonics* **2025**, *12* (1), 207–218.

(68) Sternbach, A. J.; Hinton, J.; Slusar, T.; McLeod, A. S.; Liu, M. K.; Frenzel, A.; Wagner, M.; Iraheta, R.; Keilmann, F.; Leitenstorfer, A.; Fogler, M.; Kim, H.-T.; Averitt, R. D.; Basov, D.

N. Artifact Free Time Resolved Near-Field Spectroscopy. *Opt. Express* **2017**, *25* (23), 28589–28611.

(69) Dai, S.; Ma, Q.; Yang, Y.; Rosenfeld, J.; Goldflam, M. D.; McLeod, A.; Sun, Z.; Andersen, T. I.; Fei, Z.; Liu, M.; Shao, Y.; Watanabe, K.; Taniguchi, T.; Thiemens, M.; Keilmann, F.; Jarillo-Herrero, P.; Fogler, M. M.; Basov, D. N. Efficiency of Launching Highly Confined Polaritons by Infrared Light Incident on a Hyperbolic Material. *Nano Lett.* **2017**, *17* (9), 5285–5290.

(70) Passler, N. C.; Ni, X.; Carini, G.; Chigrin, D. N.; Alù, A.; Paarmann, A. Layer-Resolved Resonance Intensity of Evanescent Polariton Modes in Anisotropic Multilayers. *Phys. Rev. B* **2023**, *107* (23), 235426.

(71) Cataldo, G.; Wollack, E. J.; Brown, A. D.; Miller, K. H. Infrared Dielectric Properties of Low-Stress Silicon Oxide. *Opt. Lett.* **2016**, *41* (7), 1364–1367.

(72) Sternbach, A. J.; Moore, S. L.; Rikhter, A.; Zhang, S.; Jing, R.; Shao, Y.; Kim, B. S. Y.; Xu, S.; Liu, S.; Edgar, J. H.; Rubio, A.; Dean, C.; Hone, J.; Fogler, M. M.; Basov, D. N. Negative Refraction in Hyperbolic Hetero-Bicrystals. *Science* **2023**, *379* (6632), 555–557.

(73) Okada, M.; Sawazaki, T.; Watanabe, K.; Taniguch, T.; Hibino, H.; Shinohara, H.; Kitaura, R. Direct Chemical Vapor Deposition Growth of $WS_2$ Atomic Layers on Hexagonal Boron Nitride. *ACS Nano* **2014**, *8* (8), 8273–8277.

(74) Nikitin, A. Y.; Alonso-González, P.; Vélez, S.; Mastel, S.; Centeno, A.; Pesquera, A.; Zurutuza, A.; Casanova, F.; Hueso, L. E.; Koppens, F. H. L.; Hillenbrand, R. Real-Space Mapping of Tailored Sheet and Edge Plasmons in Graphene Nanoresonators. *Nat. Photonics* **2016**, *10* (4), 239–243.

(75) Li, P.; Dolado, I.; Alfaro-Mozaz, F. J.; Casanova, F.; Hueso, L. E.; Liu, S.; Edgar, J. H.; Nikitin, A. Y.; Vélez, S.; Hillenbrand, R. Infrared Hyperbolic Metasurface Based on Nanostructured van der Waals Materials. *Science* **2018**, *359* (6378), 892–896.

(76) Santosh, R.; Rao, U. N.; Rao, M. J. M.; Yattirajula, S. K.; Kumar, V. The Anisotropy and Birefringence of Monolayer $WS_2$ Semiconductor. In *Micro and Nanoelectronics Devices, Circuits and Systems*; Lenka, T. R., Misra, D., Fu, L., Eds.; Springer Nature: Singapore, 2023; pp 249–255.

(77) Vyshnevyy, A. A.; Ermolaev, G. A.; Grudinin, D. V.; Voronin, K. V.; Kharichkin, I.; Mazitov, A.; Kruglov, I. A.; Yakubovsky, D. I.; Mishra, P.; Kirtaev, R. V.; Arsenin, A. V.; Novoselov, K. S.; Martin-Moreno, L.; Volkov, V. S. van der Waals Materials for Overcoming Fundamental Limitations in Photonic Integrated Circuitry. *Nano Lett.* **2023**, *23* (17), 8057–8064.

*Supporting Information for*

# Ultrafast Nano-Imaging and Optical Control of Hyperbolic Phonon Polaritons at hBN/$WS_2$ Heterojunctions

*Kazuki Kamada[1,2], Keisuke Shinokita[1,3], Fanyu Zeng[4], Ryo Kitaura[4], Kenji Watanabe[5], Takashi Taniguchi[4], Alexander Paarmann[6], Masahiro Shibuta[1,2], Takashi Kumagai[1,3*], Jun Nishida[1,3*]*

[1]Institute for Molecular Science, National Institutes of Natural Sciences, Okazaki, Aichi, 444-8585, Japan

[2]Department of Physics and Electronics, Graduate School of Engineering, Osaka Metropolitan University, 1-1, Gakuen-cho, Naka-ku, Sakai, Osaka 599-8531, Japan

[3]The Graduate University for Advanced Studies, SOKENDAI, Hayama 240-0193, Kanagawa, Japan

[4]Research Center for Materials Nanoarchitectonics, National Institute for Materials Science, 1-1 Namiki, Tsukuba 305-0044, Japan

[5]Research Center for Electronic and Optical Materials, National Institute for Materials Science, 1-1 Namiki, Tsukuba 305-0044, Japan

[6]Fritz-Haber-Institut der Max-Planck-Gesellschaft, Faradayweg 4-6, 14195 Berlin, Germany

**Table of contents**

**Note S1. Experimental details.**

**Sample preparation**

The growth of monolayer and bilayer $WS_2$ on hBN flakes was performed using a custom-built metal-organic chemical vapor deposition system. First, hBN flakes were mechanically exfoliated from a high-quality hBN crystal grown through high-temperature and high-pressure methods and then deposited onto a soda-lime glass substrate. The prepared substrate was placed on a susceptor in a load-lock chamber and transferred to the main growth chamber. The substrate-showerhead distance was set to 5 cm, and sources, including bis(tert-butylimino)bis(dimethylamino) tungsten and diethyl sulfide, were supplied through the showerhead. The sources are in liquid form and can be easily supplied by bubbling Ar buffer gas through them. Typically, the substrate temperature was increased to 973 K while flowing the sulfur source, and the tungsten source was supplied for 10 minutes. After growing $WS_2$, the sulfur source flow was maintained until the temperature dropped below 773 K to prevent the formation of sulfur vacancies. All growth procedures, including valve operations, flow, and temperature control, were carried out with LabVIEW-based software.

The hBN/$WS_2$ heterostructure was prepared using the polymer stamp dry-transfer technique. Mechanically exfoliated $WS_2$ and hBN on a $SiO_2$/Si substrate were picked up and stacked using a polyvinyl chloride (PVC) sheet on a polydimethylsiloxane (PDMS) lens.[1]

**Laser system.**

Most of the output from a Yb:KGW femtosecond oscillator (FLINT, Light Conversion; central wavelength 1030 nm, pulse duration ~ 150 fs, repetition rate ~ 76 MHz, pulse energy ~ 0.12 μJ; total power ~ 9 W) is used to pump an optical parametric oscillator (OPO, Levante fs, APE GmbH), which produces signal and idler pulses. The OPO incorporates an active stabilization system employing piezoelectric actuators, enabling steady output power and spectrum as well as automatic wavelength tuning. Difference-frequency generation (DFG) between the signal and idler beams is carried out using a HarmoniXX DFG module (APE GmbH), generating wavelength-tunable mid-infrared pulses covering 1000–2000 $cm^{-1}$. These pulses exhibit a spectral full width at half maximum of roughly 150 $cm^{-1}$ and an output power in the range of 5–30 mW. During the experiment performed in the manuscript, the center frequency of the mid-infrared pulse was set to ~1500 $cm^{-1}$. A 515 nm pump beam is obtained by second-harmonic generation of a portion of the Yb:KGW oscillator output using a 1 mm-thick BBO crystal (CASTECH).

**Narrowband resolved polariton imaging.**

A commercial nano-FTIR system (neaSNOM, Neaspec GmbH) was employed, incorporating an atomic force microscope (AFM), an asymmetric Michelson interferometer with an integrated reference arm, and a mercury–cadmium–telluride (MCT) detector (J19D12-M204-R250U-60-WE, Teledyne Judson Technologies). For narrowband detection, optical grating (53067BK01-933R, Richardson Gratings) was placed before the MCT detector, which allows imaging of phonon polaritons. The MCT output was passed through a high-pass filter (EF507, Thorlabs) to remove the DC and first-harmonic components. The filtered signal was subsequently amplified (DHPVA-101, FEMTO Messtechnik GmbH) and processed using an external digital lock-in amplifier (HF2LI, Zurich Instruments).

The AFM was operated in tapping mode, with the lock-in amplifier referenced to the tip oscillation frequency ($\omega_{\mathrm{AFM}}$). The near-field signal was demodulated at higher harmonics of the tapping frequency ($n\omega_{\mathrm{AFM}}$) to isolate the near-field contribution. A Pt/Ir-coated silicon probe (ARROW-NCPt, NanoWorld; resonant frequency ~ 285 kHz, spring constant ~ 42 N/m) was employed with an additional Au coating by sputter deposition to enhance tip reflectivity.

For ultrafast infrared *s*-SNOM measurements, the pump beam was modulated at $\Omega_{\mathrm{chop}}$ ~ 10 kHz using a mechanical chopper (MC2000B, Thorlabs). The chopper reference signal was fed into the lock-in amplifier, which generated the sideband frequencies ($n\omega_{\mathrm{AFM}} \pm \Omega_{\mathrm{chop}}$) to enable selective detection of pump–probe near-field responses. To suppress background interference, the scattered near-field signal was interfered with a temporally controlled reference pulse, and measurements were conducted at multiple optical phase conditions.

### Numerical simulations

Numerical simulations using finite element method (JCMwave) were performed to calculate field distribution in two configurations of $WS_2$ (1.5 nm)/hBN (120 nm)/$SiO_2$ (2000 nm) and hBN (38 nm)/$WS_2$ (376 nm)/$SiO_2$ (1624 nm). Dipole source modeling AFM tip is located at 300 nm above hBN surface and scanned across the $WS_2$/hBN heterostructure. Geometry used for the simulation is shown in Figure 5 in the main text. For hBN/$WS_2$/$SiO_2$ structures, we did not consider Si layer below $SiO_2$, because polaritonic field was sufficiently decayed within $WS_2$ and $SiO_2$ layer.

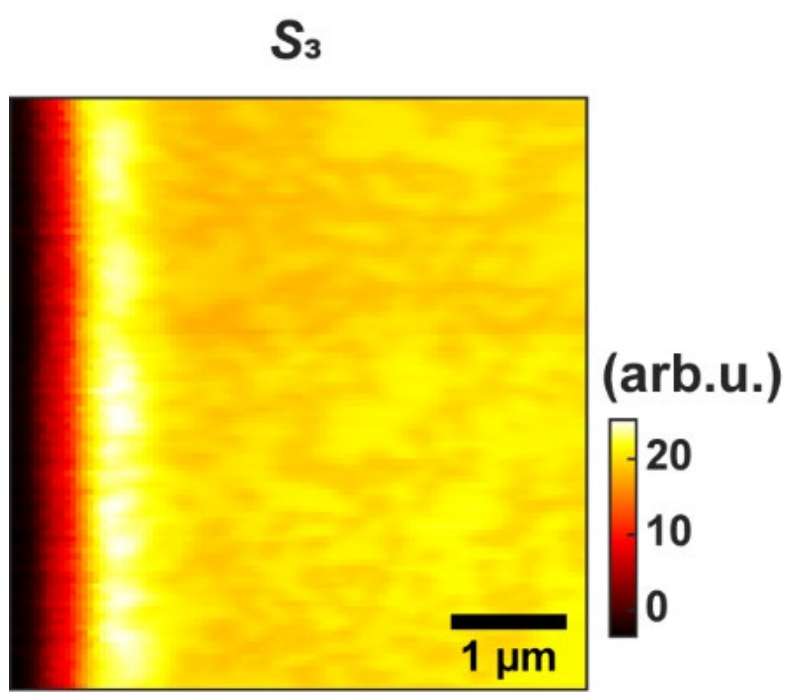

**Figure S1.** Near-field imaging in broadband detection without a grating. A near-field image at the edge of hBN in broadband detection without the use of an optical grating. The center frequency of the incident mid-infrared pulse was 1490 $cm^{-1}$ with the spectral bandwidth of ~150 $cm^{-1}$.

**Note S2 Edge-launched polaritons.**

In our grating-coupled detection scheme, we identified fringe patterns originating not only from the interference between tip-launched and edge-reflected polaritons but also from directly edge-launched polaritons. Figures S2a and S2b display near-field images of exfoliated $WS_2$ on hBN in the ground state (a) and transient state (b), where the $WS_2$ layer (2.5 nm thick) is homogeneously distributed over the hBN substrate (125 nm thick). Pronounced fringe patterns are evident near the hBN edge. Figure S2c presents the corresponding fringe profiles of $S_3$ and $\Delta S_2$, revealing oscillations with different periodicities in both signals. These oscillations are attributed to tip-launched ($\lambda_p/2$) and edge-launched ($\lambda_p$) polariton modes. The edge-launched polaritons propagate as plane waves, in contrast to the circular wavefronts of tip-launched polaritons, leading to longer propagation distances owing to the absence of geometrical damping. To determine the polariton wavelength, the fringe patterns were subjected to Fourier transformation, and the spectral amplitudes of $S_3$ and $\Delta S_2$ ($|S_3|$ and $|\Delta S_2|$) are plotted in Figure S2d, where two distinct peaks corresponding to different origins of excitation are clearly discernible. The dispersion relation extracted from the polariton wavelengths of these peaks at varying frequencies in $|S_3|$ demonstrates good agreement with theoretical predictions obtained via the transfer matrix method for the $WS_2$ (2.5 nm)/hBN (125 nm)/$SiO_2$ heterostructure (Figure S2e).

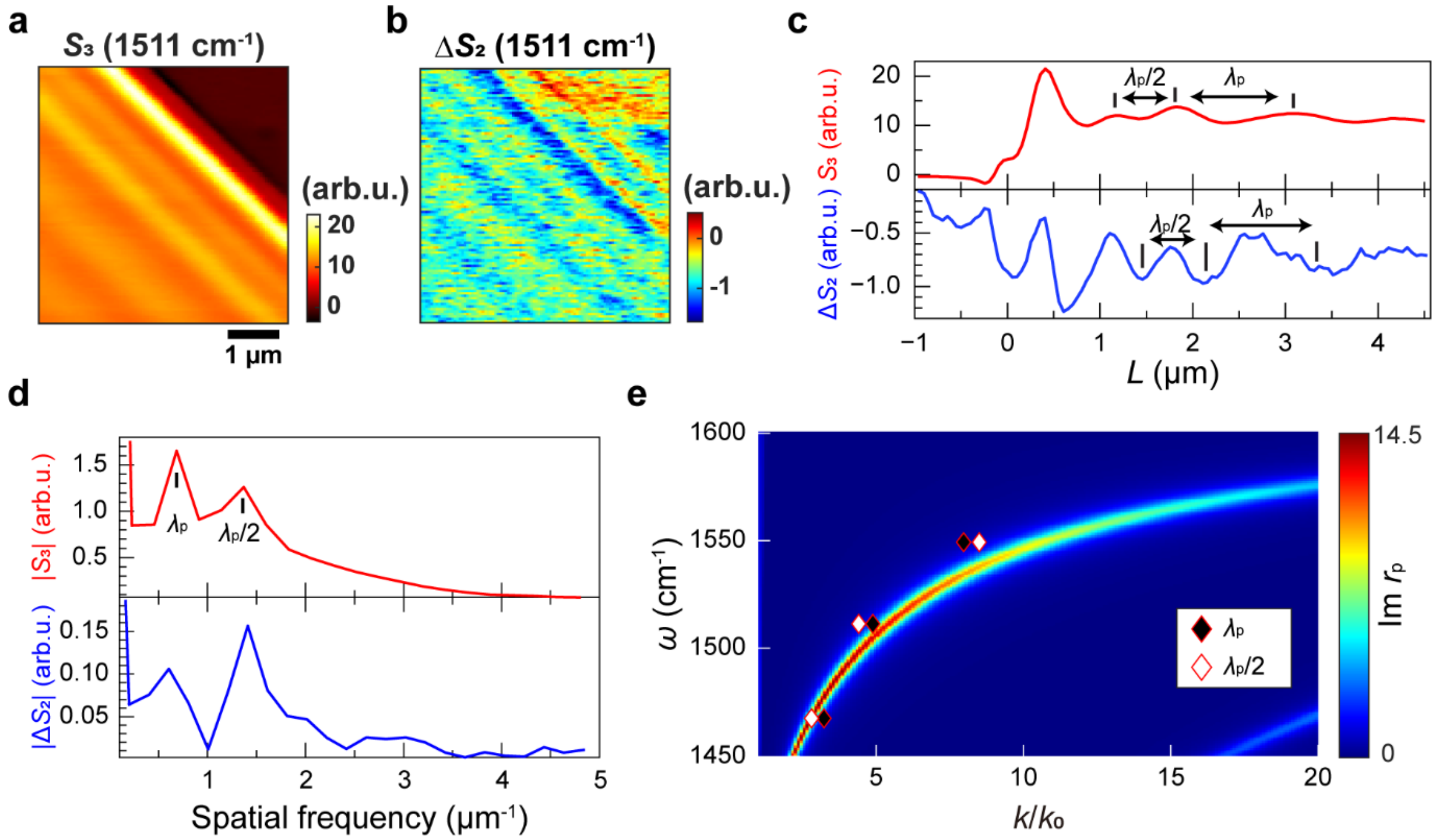

**Figure S2.** Near-field images of edge-launched polaritons. (a–b) Near-field images of exfoliated $WS_2$/hBN heterostructures at a frequency of 1511 cm$^{-1}$, corresponding to the ground state (a) and transient state (b) of $WS_2$. The scalar bar represents 1 μm in these images. (c) Fringe profiles of $S_3$ (red line) and $\Delta S_2$ (blue line) extracted from (a) and (b). Tip- and edge-launched components are indicated by black arrows, corresponding to $\lambda_p/2$ and $\lambda_p$, respectively. (d) Fourier-transformed amplitudes of $S_3$ ($|S_3|$, red line) and $\Delta S_2$ ($|\Delta S_2|$, blue line), with $\lambda_p/2$ and $\lambda_p$ marked by black lines. (e) Diamonds: dispersion relation of $S_3$ derived from the $\lambda_p/2$ (white) and $\lambda_p$ (black) components of $|S_3|$ at three different frequencies. Colormap: calculated imaginary part of the complex reflection coefficient (Im $r_p$) obtained via the transfer matrix method for the $WS_2$ (2.5 nm)/hBN (125 nm)/$SiO_2$ structure. The scale bar represents 1 μm in all panels.

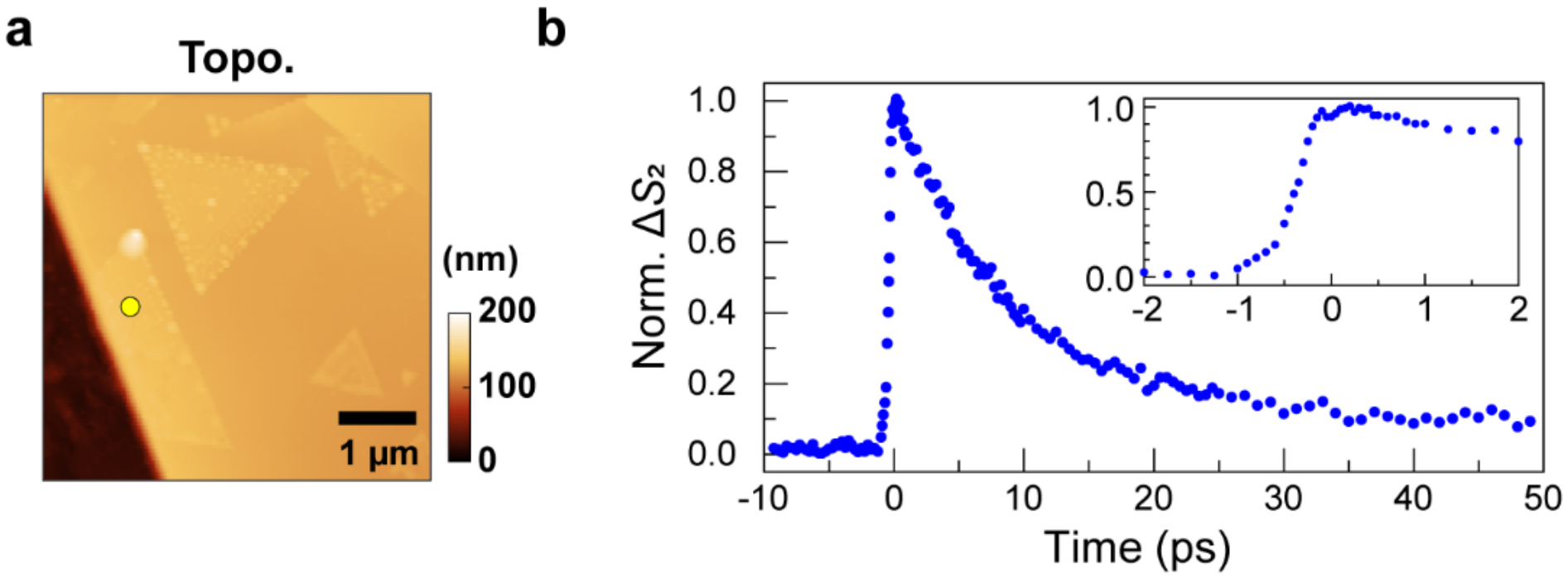

**Figure S3.** Near-field pump–probe time trace acquired with the narrowband detection scheme. (a) AFM topography, which is identical to that shown in Figure S5a. A yellow dot indicates the location where the near-field pump–probe measurement was performed. (b) Near-field pump-probe time trace. The inset shows a magnified view of the interval from -2 ps to 2 ps.

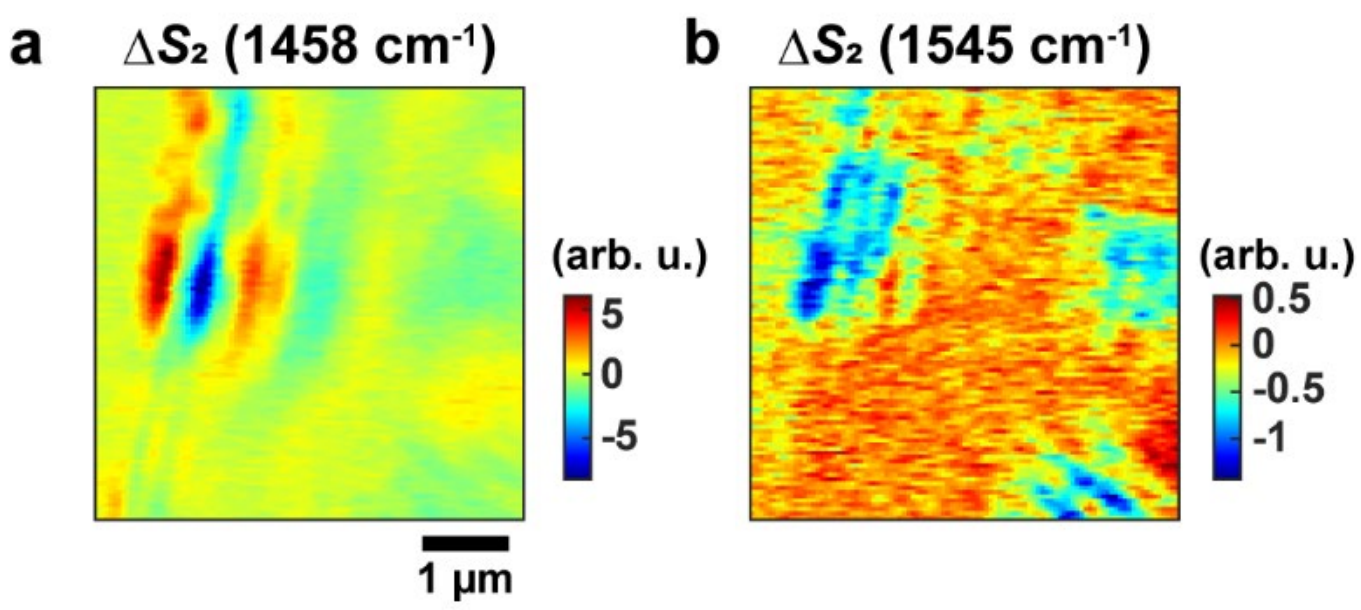

**Figure S4.** Ultrafast pump–probe near-field imaging of CVD-grown $WS_2$/hBN/$SiO_2$ heterostructures: different probe frequencies. (a‑b) Transient near-field images at the probe frequency of (a) 1458 $cm^{-1}$ and (b) 1545 $cm^{-1}$. The scale bar represents 1 μm in all images.

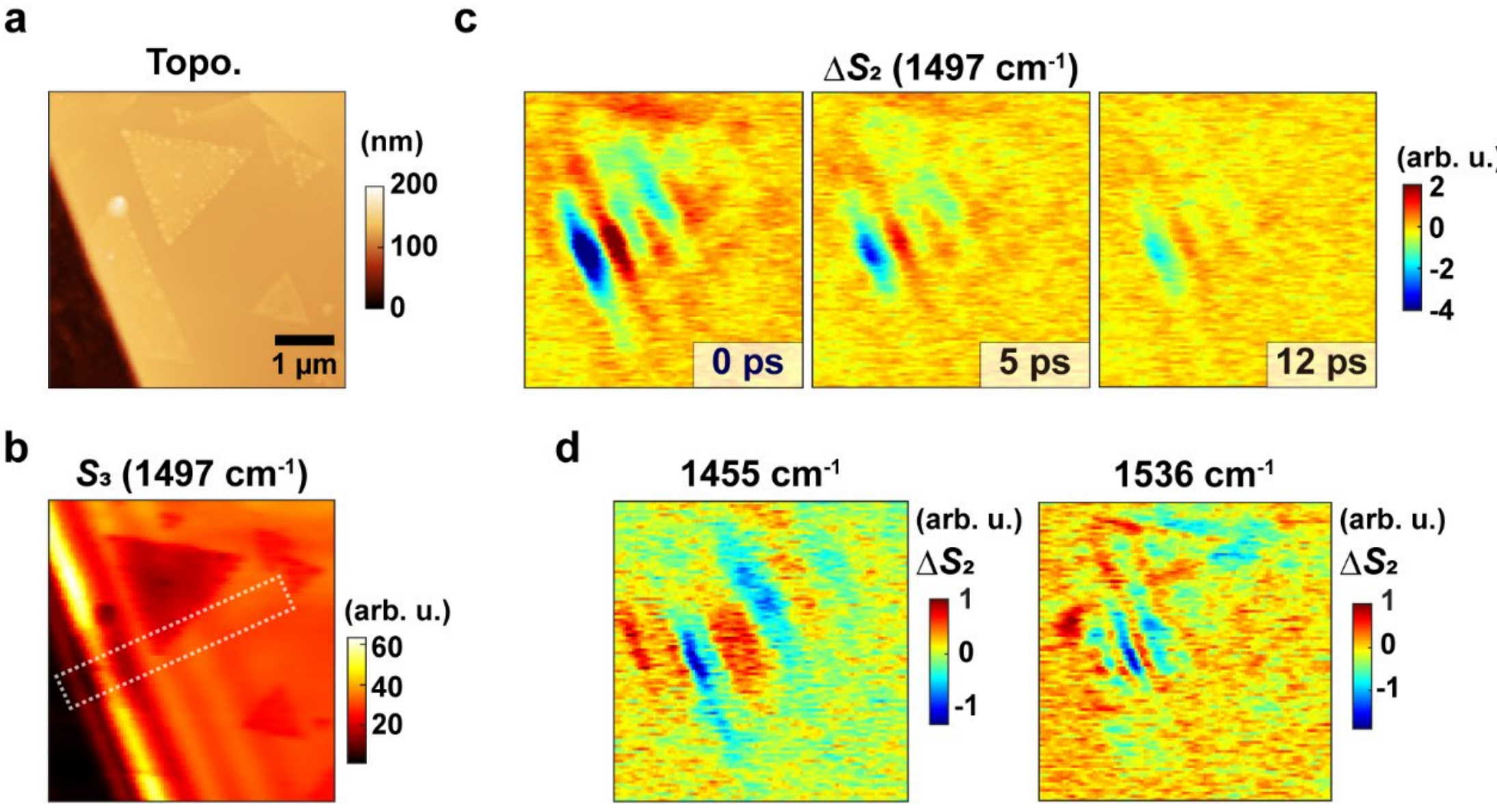

**Figure S5.** Ultrafast pump–probe near-field imaging of CVD-grown $WS_2$/hBN/$SiO_2$ heterostructures: another data set. (a) AFM topography of the $WS_2$/hBN/$SiO_2$ structure which is the same topography as Figure S3a. (b) A near-field ground-state image at a frequency of 1497 cm$^{-1}$. (c–d) Near-field pump–probe images at different probe frequencies of (c) 1497 cm$^{-1}$ at different delay times of 0, 5, 12 ps, and (d) 0 ps at different frequencies of 1455 cm$^{-1}$, 1536 cm$^{-1}$. The scale bar represents 1 μm in all images.

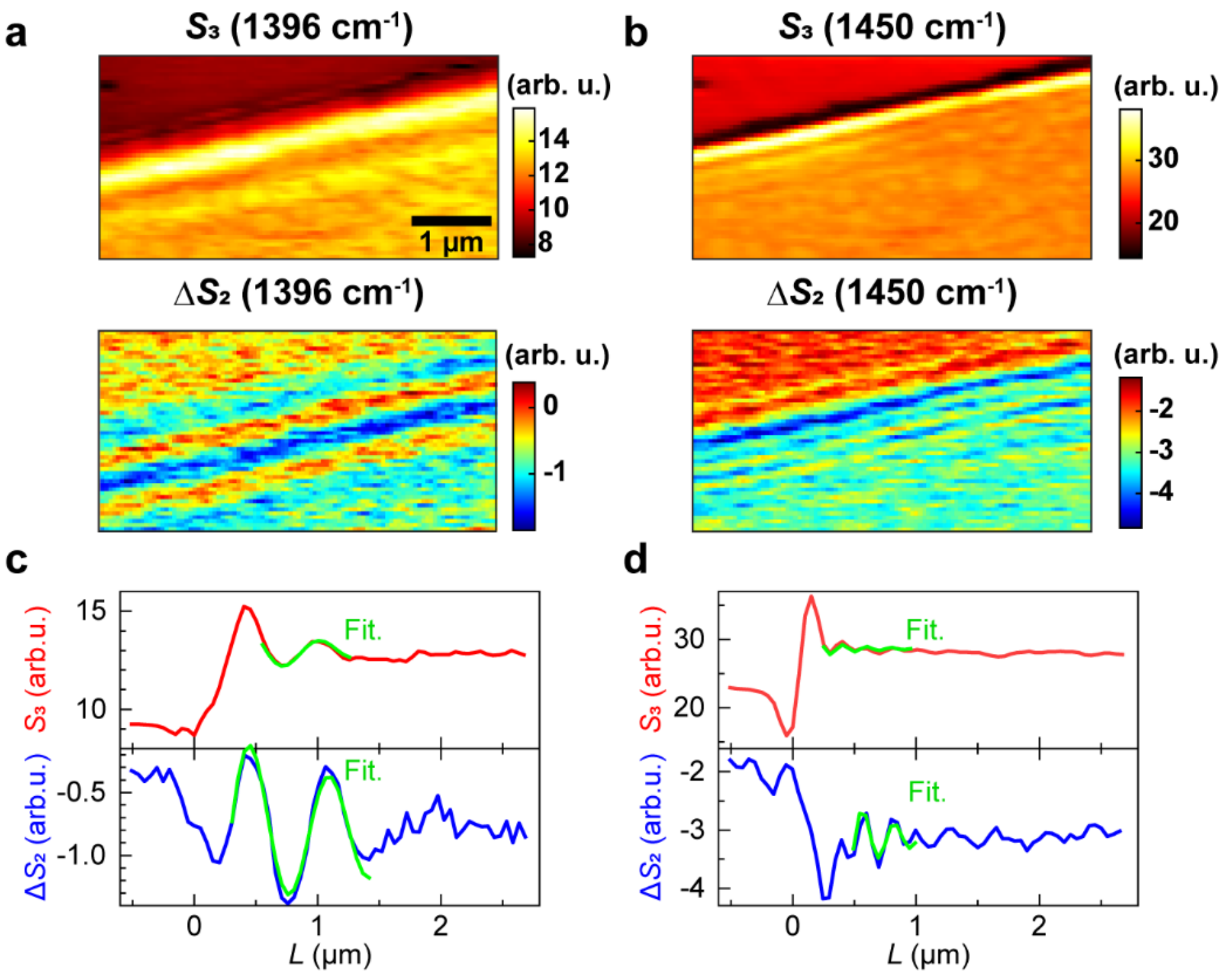

**Figure S6.** Frequency dependence in hBN (38 nm)/$WS_2$ (376 nm) /$SiO_2$. (a–b) Near-field (top) ground-state and (bottom) pump–probe images at the frequency of (a) 1396 $cm^{-1}$ and (b) 1450 $cm^{-1}$. (c–d) Fringe profiles $S_3$ (red line) and $\Delta S_2$ (blue line) at the frequency of (c) 1396 $cm^{-1}$ and (d) 1450 $cm^{-1}$. Green line shows the fitting. Scale bar represents 1 μm in all images.

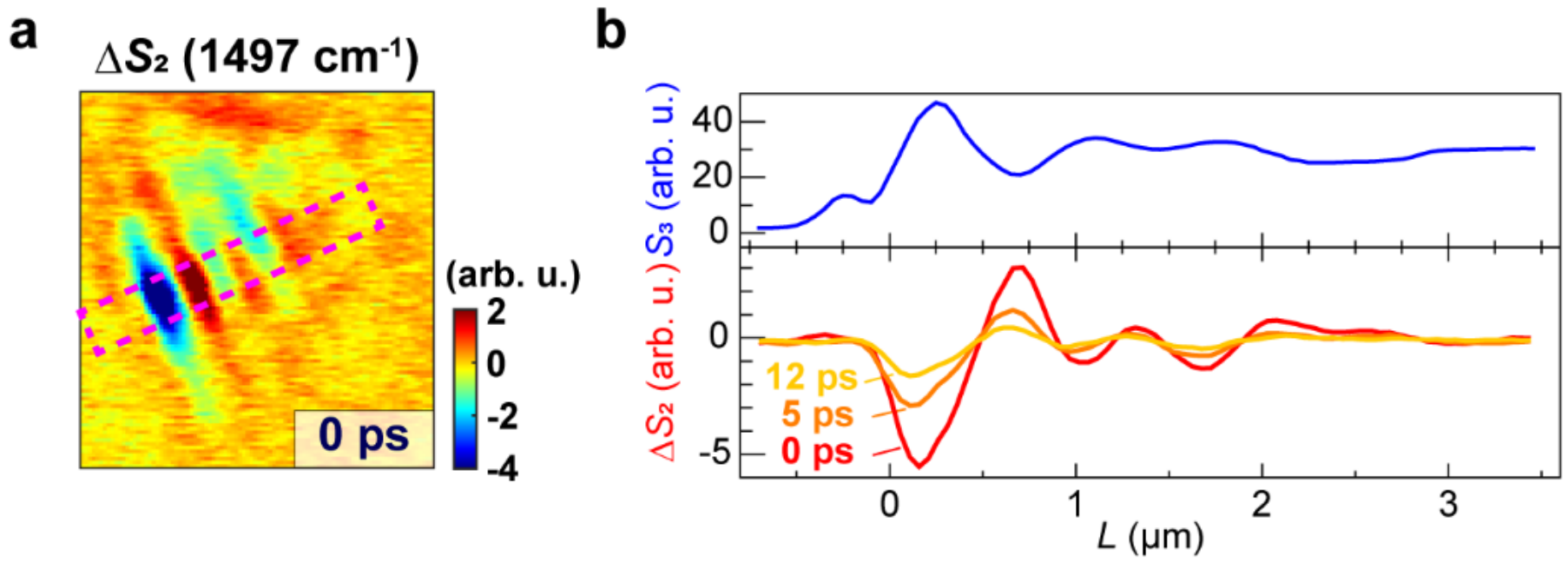

**Figure S7.** Line profiles extracted from the data in Figure S5. (a) Near-field transient image identical to that shown in Supporting Information Figure S5c. We extracted line profiles within the magenta dashed rectangle. (b) Line profiles of $S_3$ and $\Delta S_2$ at 0, 5, 12 ps. Because of the uncertainty of hBN thickness in Figure 3b and 3c, we instead used an additional dataset (Supporting Information Figure S5) with well-defined hBN thickness for comparison with numerical calculations in Figure 5.

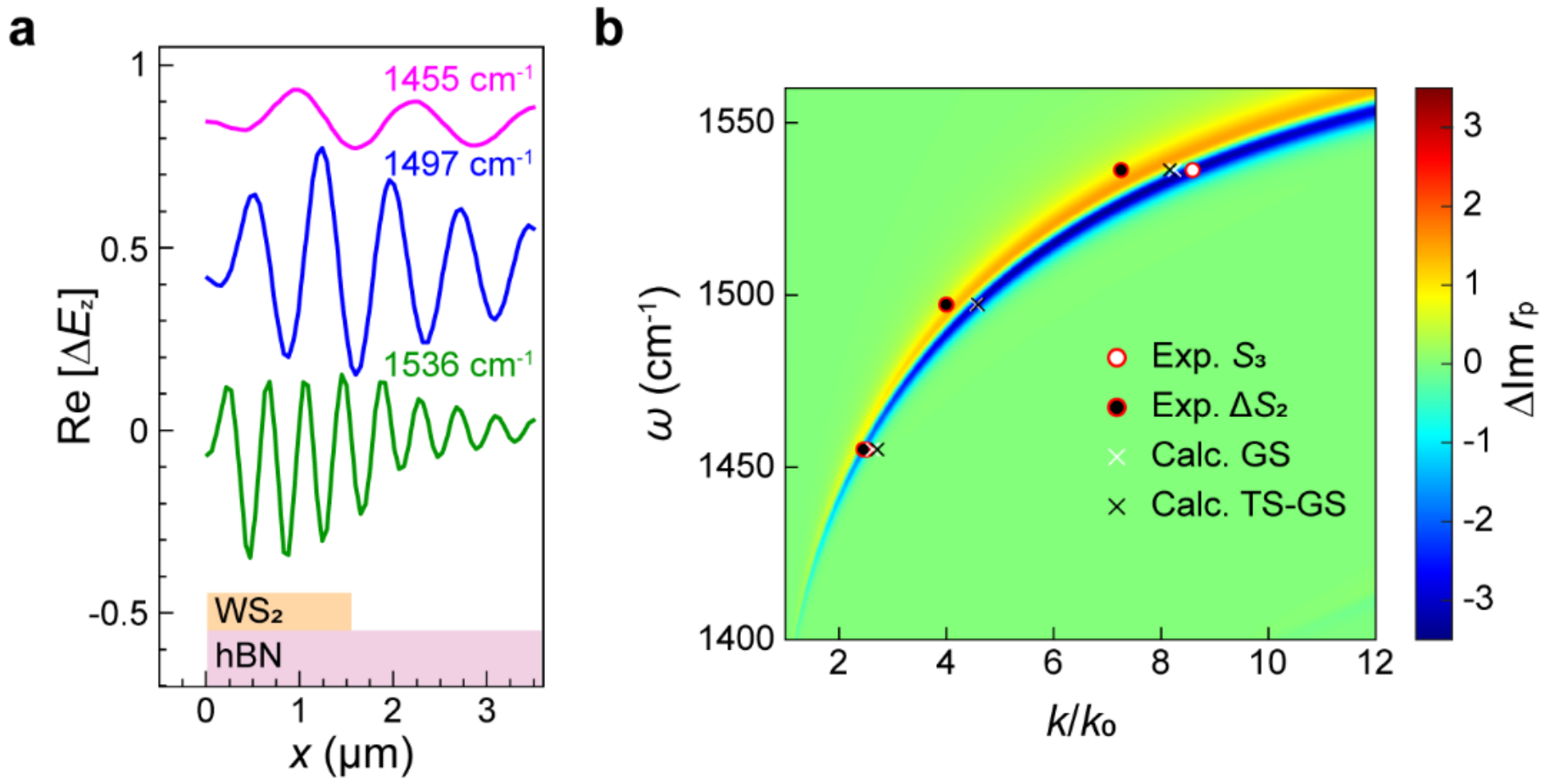

**Figure S8.** Dipole-frequency dependence of $\mathrm{Re}[\Delta E_Z]$. (a) $\mathrm{Re}[\Delta E_Z]$ calculated at different dipole frequencies: 1455 cm$^{-1}$ (pink line), 1497 cm$^{-1}$ (blue line), and 1536 cm$^{-1}$ (green line). The spatial configurations of the $WS_2$ and hBN layers are indicated at the bottom of the figure. (b) Two-dimensional plot of $\Delta\mathrm{Im}\ r_p$, defined as the difference between $\mathrm{Im}\ r_p$ in the transient and ground states. Circular markers represent experimental data obtained from the profiles in Figure 3d, while cross markers denote numerical simulation results presented in (a).

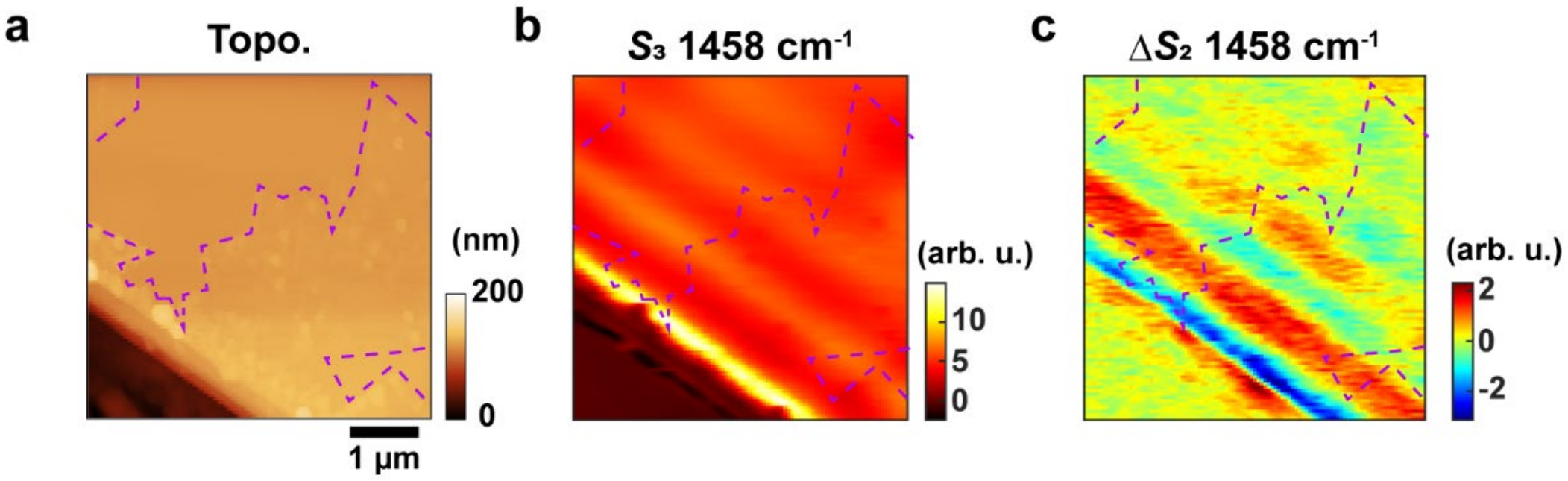

**Figure S9.** Fringe oscillations extending beyond the $WS_2$ region in near-field pump–probe imaging. (a) AFM topography of the CVD-grown $WS_2$/hBN heterostructure. (b–c) Near-field (b) ground-state and (c) pump–probe image of $WS_2$ at the ground state acquired at a probe frequency of 1458 cm$^{-1}$. The $WS_2$ region is outlined by a purple line in all images. The scale bar represents 1 μm in all panels.

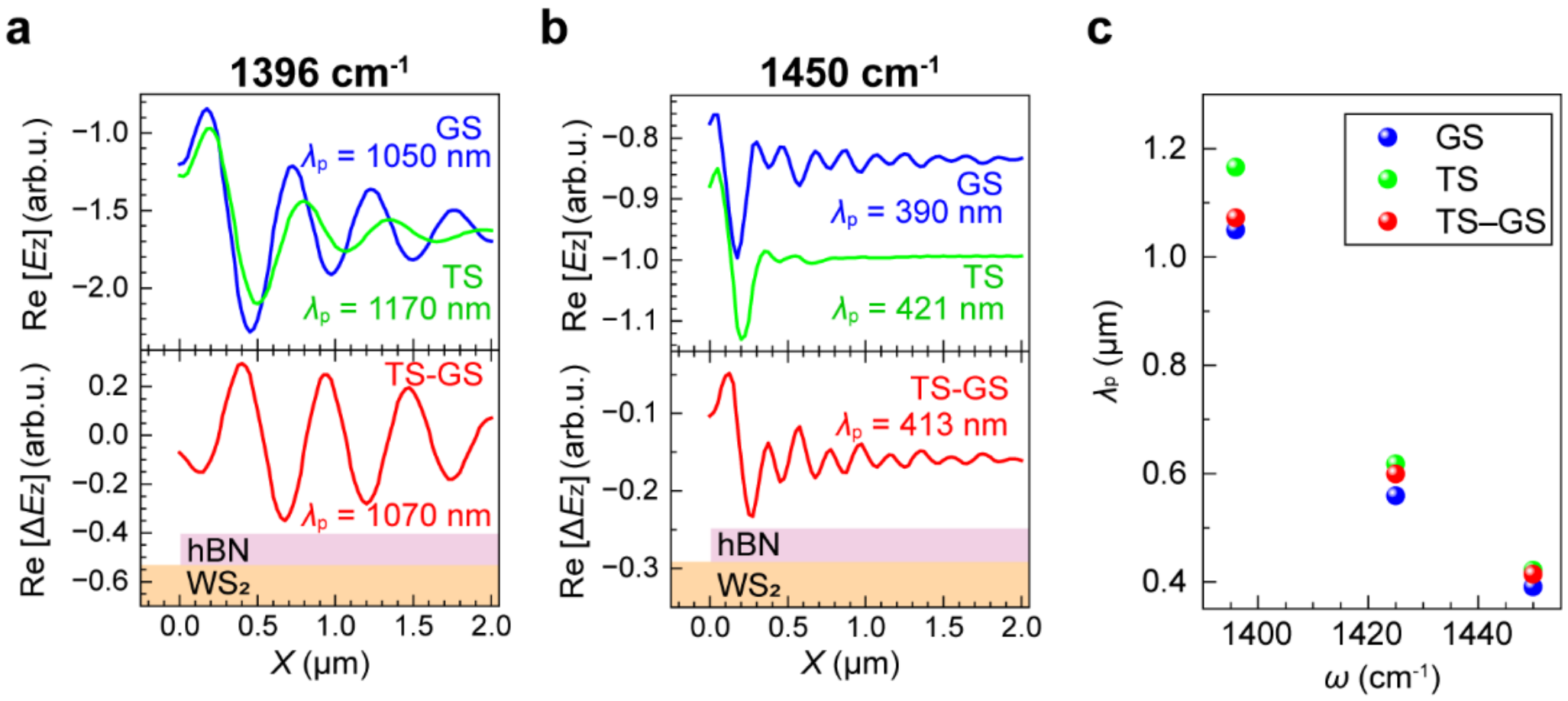

**Figure S10.** Frequency dependence of the numerical simulation in the hBN/$WS_2$/$SiO_2$ configuration. (a–b) $\mathrm{Re}\left[E_Z^{\mathrm{GS}}\right]$ (blue line) and $\mathrm{Re}\left[E_Z^{\mathrm{TS}}\right]$ (green line) of $WS_2$ (top) and $\mathrm{Re}[\Delta E_Z]$ is shown by red line (TS–GS) (bottom) at different dipole frequencies of (a) 1396 $cm^{-1}$ and (b) 1450 $cm^{-1}$. The schematics of (a) $WS_2$/hBN and (b) hBN/$WS_2$ used in the calculation are shown at the bottom of each figure. (c) $\lambda_p$ of GS (red circle), TS (green circle) and TS–GS (blue circle) as a function of dipole source frequency ($\omega$).

**Note S3 Pump-fluence dependence**

Our results suggest that HPhPs can be controlled by the carrier density in $WS_2$. To verify this, we investigated the pump-fluence dependence of transient near-field imaging of HPhPs. Figure S11a shows transient near-field images at different pump fluence of 8.0, 19, 23, and 32 μJ/cm$^2$ acquired from the same region shown in Figure 4. Corresponding fringe profiles extracted from these images are displayed in Figure S11b. A distinct phase shift near the sample edge ($L$ = 0 μm) becomes evident at higher pump fluence, exhibiting a displacement of 60 nm between 8.0 and 32 μJ/cm$^2$. The magnitude of this shift decreases with increasing distance from the edge, amounting to 20 nm at farther positions. In contrast, within the experimental uncertainty, we did not observe a clear dependency of the polariton wavelength on the pump fluence (Figure S11c). To elucidate the origin of the experimental behavior, we performed numerical simulations examining the dependence of the fringe profiles on the absolute value of $\Delta\varepsilon$ ($|\Delta\varepsilon|$), because higher pump fluence corresponds to an increased carrier density in $WS_2$ and thus a stronger Drude-type perturbation in the excited state (Figure S11d). The upper and lower panels depict $\mathrm{Re}[E_Z]$ for GS (black) and TS (varying colors) and $\mathrm{Re}[\Delta E_Z]$, respectively. As $|\Delta\varepsilon|$ increases, the phase of $\mathrm{Re}[\Delta E_Z]$ shifts, and the magnitude of this shift becomes larger closer to the hBN edge, consistent with our experimental observations. This phase shift originates from the change in polariton wavelength in the TS (Figure S11d, top). However, because $\mathrm{Re}\left[E_Z^{\mathrm{TS}}\right]$ decays more rapidly at larger $|\Delta\varepsilon|$, the differential signal $\mathrm{Re}[\Delta E_Z]$ becomes dominated by $\mathrm{Re}\left[E_Z^{\mathrm{GS}}\right]$ at positions farther from the edge. Consequently, the phase of $\mathrm{Re}[\Delta E_Z]$ exhibits only a small shift at these more distant positions. Regarding the polariton wavelength, $\lambda_{\mathrm{p}}^{\mathrm{TS-GS}}$ increases with larger $|\Delta\varepsilon|$ (Figure S11e). Yet, the relatively small change in the polariton wavelength, amounting to only ~40 nm of additional elongation from the lowest to the highest fluence, would be overwhelmed by the experimental uncertainty in the extracted wavelength. This explains the lack of the clear fluence dependence of the polaritonic wavelength on the experimentally extracted polaritonic wavelength.

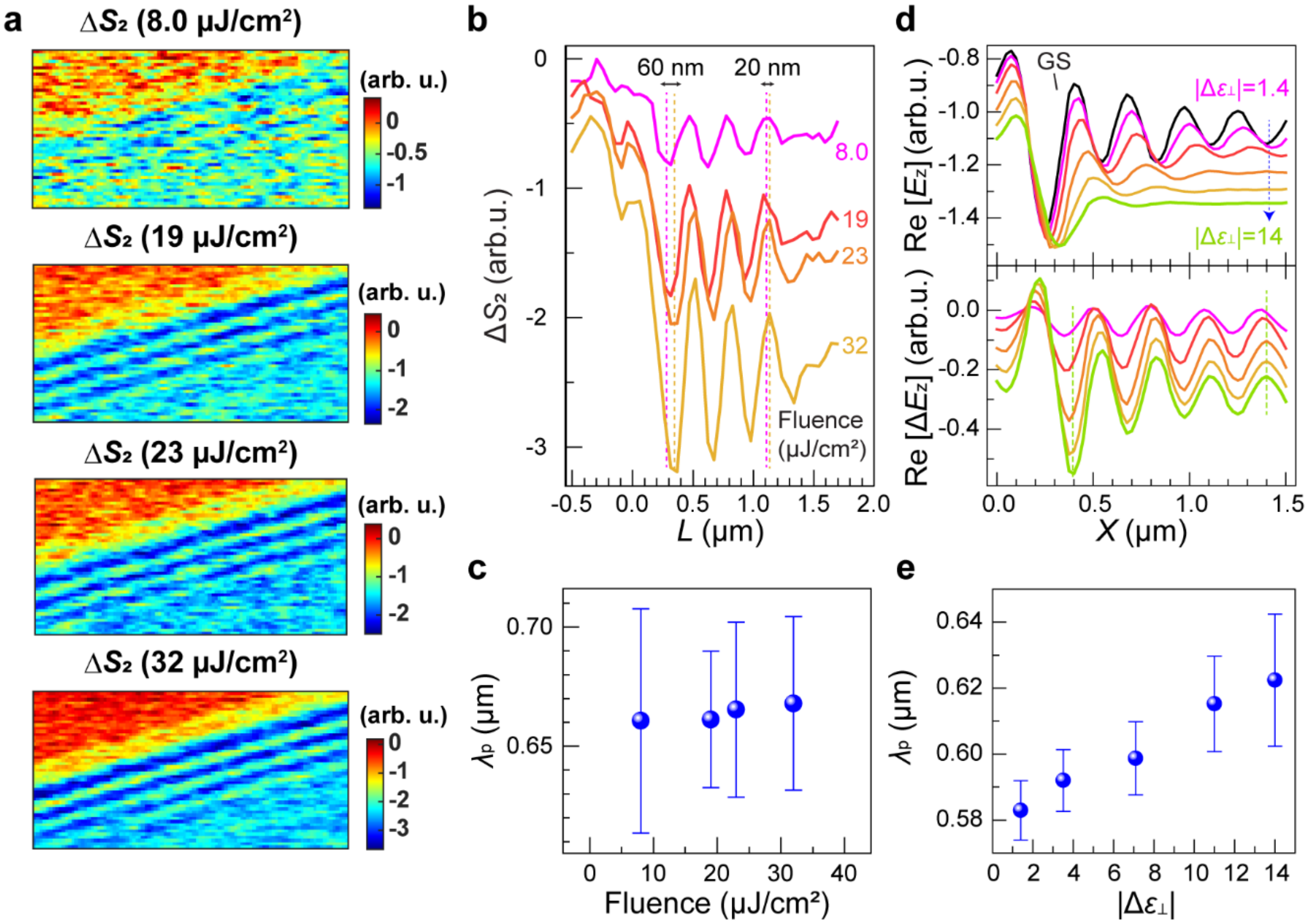

**Figure S11.** Pump-fluence dependence and numerical simulation results. (a) Near-field pump–probe images of $\Delta S_2$ at pump fluences of 8.0, 19, 23, and 32 μJ/cm² acquired from the same region shown in Figure 4. (b) Fringe profiles corresponding to (a). Dashed lines indicate the local peaks at 8.0 and 32 μJ/cm² to quantify the shift. (c) Polariton wavelength $\lambda_p$ extracted from (b). Fitting was performed over the range of 0.16–0.78 μm. (d) Numerically calculated $\mathrm{Re}[E_Z]$ at the ground state (GS, black line), transient state (TS), and $\mathrm{Re}[\Delta E_Z]$. The magnitude of $|\Delta\varepsilon_\perp|$ was varied from 1.4 to 10. (e) Polariton wavelength $\lambda_p$ extracted from $\mathrm{Re}[\Delta E_Z]$ in (c) as a function of $|\Delta\varepsilon_\perp|$. Fitting was performed over the range of 0.2–0.75 μm. Error bars in (c) and (e) represent fitting uncertainties.

**Note S4 The effect of dielectric function of $WS_2$ on the simulation results.**

In the main text, the dielectric function of $WS_2$ in the excited state is represented as $\varepsilon_\perp + \Delta\varepsilon_\perp$ and $\varepsilon_\parallel + \Delta\varepsilon_\parallel$, where $\Delta\varepsilon_\perp = -5 + 5i$ and $\Delta\varepsilon_\parallel = -0.5 + 0.5i$ were used. $\Delta\varepsilon_\parallel$ values were estimated in our previous work where the excitation fluence is comparable to the present study.[2] For the modulated in-plane dielectric response $\Delta\varepsilon_\perp$, a previous report on a thick layer of $WSe_2$ suggests $\Delta\varepsilon_\perp$ is 4 to 5 times larger than $\Delta\varepsilon_\parallel$.[3] We used these insights as a starting point and then subsequently varied the dielectric function as a tunable parameter in our simulations, elucidating how phase and amplitude affect the modulation of the electric field. First, we calculated $\mathrm{Re}[\Delta E_Z]$ while varying the phase angle of $\Delta\varepsilon$ from $\pi$ to $\pi/2$ (Figure S12a). The results reveal that the relative phase between $\mathrm{Re}[\Delta E_Z]$ and $\mathrm{Re}\left[E_Z^{\mathrm{GS}}\right]$ is governed by the phase angle of $\Delta\varepsilon$, exhibiting a phase shift of approximately $\pi/2$. This observation implies that the relative phase between $\mathrm{Re}[\Delta E_Z]$ and $\mathrm{Re}\left[E_Z^{\mathrm{GS}}\right]$ provides direct insight into the phase angle of $\Delta\varepsilon$. Subsequently, we independently varied the amplitudes of $\Delta\varepsilon_\perp$ and $\Delta\varepsilon_\parallel$, as illustrated in Figure S12b. Variations in $\Delta\varepsilon_\perp$ produced a pronounced modulation in $\mathrm{Re}[\Delta E_Z]$, whereas changes in $\Delta\varepsilon_\parallel$ yielded negligible effects.

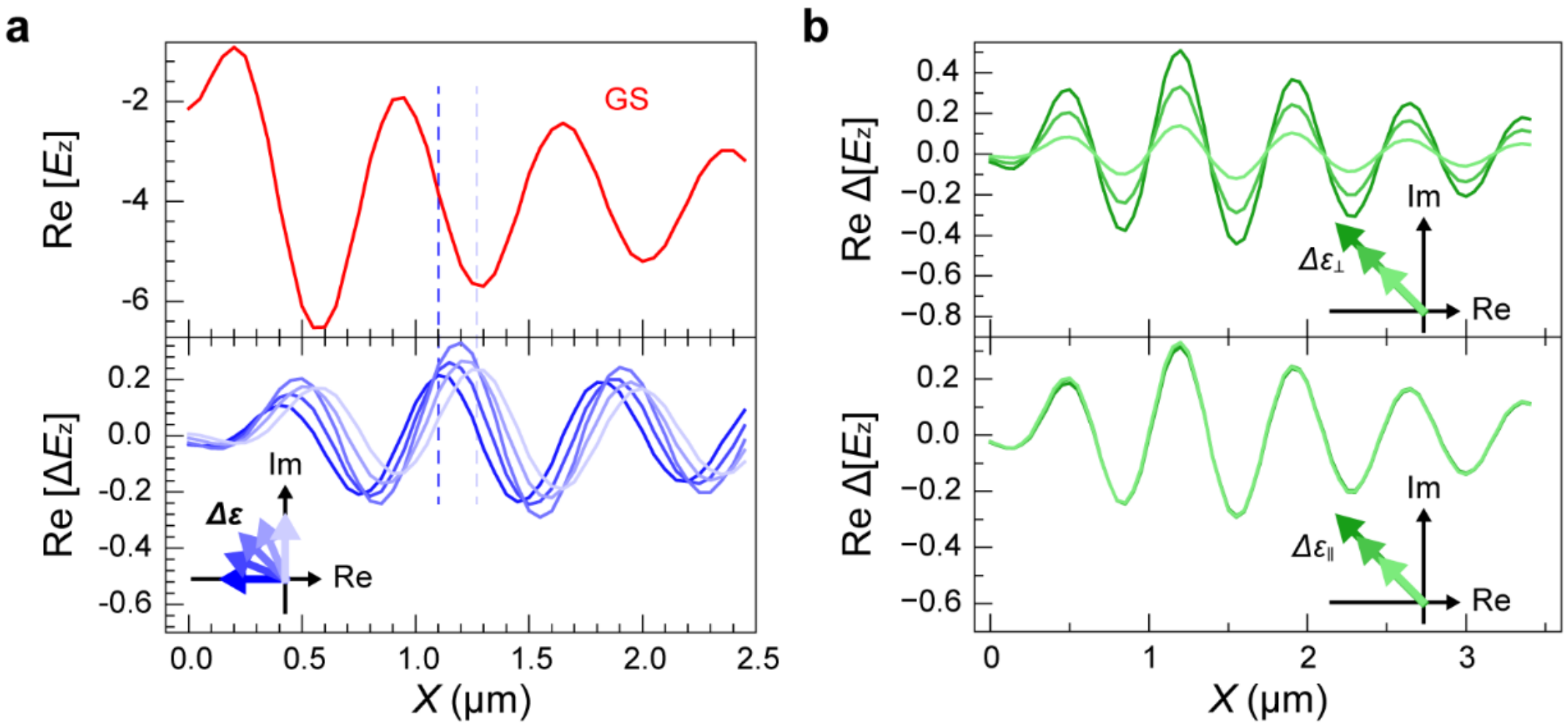

**Figure S12.** Dependence of Re[$\Delta E_Z$]on the phase and amplitude of $\Delta\varepsilon$. (a) Real part of $E_z$ (Re[$E_Z$]) at the ground state (GS) (top) and Re[$\Delta E_Z$] calculated by varying the phase angle of $\Delta\varepsilon$ from π to π/2 (bottom). The corresponding dielectric perturbations are [$\Delta\varepsilon_\perp$, $\Delta\varepsilon_\parallel$] = [–5, –0.5], [–5 + 2.5*i*, –0.5 + 0.25*i*], [–5 + 5*i*, –0.5 + 0.5*i*], [–2.5 + 5*i*, –0.25 + 0.25*i*], and [5*i*, 0.5*i*]. (b) Re[$\Delta E_Z$] as a function of the amplitude of $\Delta\varepsilon_\perp$ (top; $\Delta\varepsilon_\perp$ = –2 + 2*i*, –5 + 5*i*, –8 + 8*i*, with $\Delta\varepsilon_\parallel$ fixed at –0.5 + 0.5*i*) and $\Delta\varepsilon_\parallel$ (bottom; $\Delta\varepsilon_\parallel$ = –0.5 + 0.5*i*, –3 + 3*i*, –7 + 7*i*, with $\Delta\varepsilon_\perp$ fixed at –5 + 5*i*).

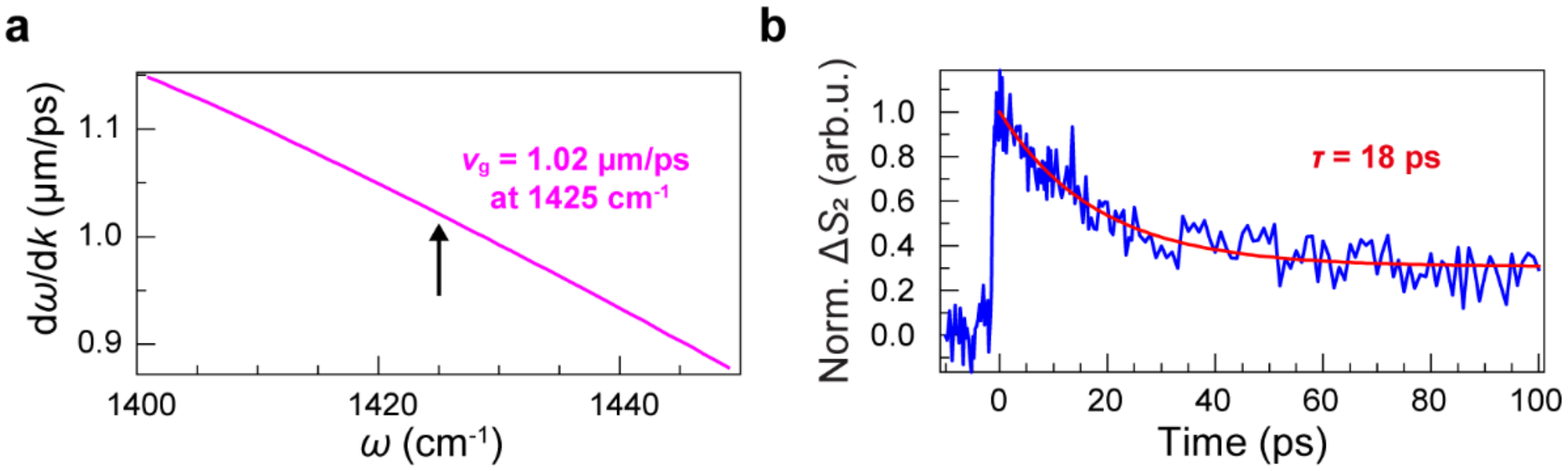

**Figure S13.** Calculated group velocities and near-field pump–probe time trace in the hBN (38 nm)/$WS_2$ (376 nm)/$SiO_2$ structure. (a) Calculated group velocities ($v_g$) derived from Im $r_p$. The group velocity was determined to be 1.02 μm/ps at 1425 $cm^{-1}$. (b) Near-field pump–probe time trace acquired under narrowband detection at 1425 $cm^{-1}$. The tip was positioned sufficiently far from the edge.

**References**

(1) Wakafuji, Y.; Onodera, M.; Masubuchi, S.; Moriya, R.; Zhang, Y.; Watanabe, K.; Taniguchi, T.; Machida, T. Evaluation of Polyvinyl Chloride Adhesion to 2D Crystal Flakes. *Npj 2D Mater. Appl.* **2022**, *6* (1), 44.

(2) Wang, Y.; Nishida, J.; Nakamoto, K.; Yang, X.; Sakuma, Y.; Zhang, W.; Endo, T.; Miyata, Y.; Kumagai, T. Ultrafast Nano-Imaging of Spatially Modulated Many-Body Dynamics in CVD-Grown Monolayer $WS_2$. *ACS Photonics* **2025**, *12* (1), 207–218.

(3) Sternbach, A. J.; Chae, S. H.; Latini, S.; Rikhter, A. A.; Shao, Y.; Li, B.; Rhodes, D.; Kim, B.; Schuck, P. J.; Xu, X. Programmable Hyperbolic Polaritons in van der Waals Semiconductors. *Science*, **2021**, *371* (6529), 617–620.